\documentclass[iop,revtex4]{emulateapj}

\usepackage{lscape}

\newcommand{\HII}{{\ion{H}{2}}}

\newcommand{\HeI}{{\ion{He}{1}}}
\newcommand{\HI}{{\ion{H}{1}}}
\newcommand{\SI}{{\ion{S}{1}}}
\newcommand{\SiIIIf}{[{\ion{Si}{3}}]}
\newcommand{\SiIII}{{\ion{Si}{3}}]}
\newcommand{\NIII}{{\ion{N}{3}}]}
\newcommand{\NI}{[{\ion{N}{1}}]}
\newcommand{\NeIV}{[{\ion{Ne}{4}}]}

\newcommand{\ArIV}{[{\ion{Ar}{4}}]}
\newcommand{\ClIII}{[{\ion{Cl}{3}}]}

\newcommand{\OII}{[{\ion{O}{2}}]}
\newcommand{\AlII}{[{\ion{Al}{2}}]}
\newcommand{\AlIIint}{{\ion{Al}{2}}]}
\newcommand{\AlIIres}{{\ion{Al}{2}}}
\newcommand{\CII}{[{\ion{C}{2}}]}
\newcommand{\OI}{[{\ion{O}{1}}]}

\newcommand{\OV}{[{\ion{O}{5}}]}

\newcommand{\CI}{[{\ion{C}{1}}]}

\def\ratioR23{([\ion{O}{2}]~$\lambda$3726 +[\ion{O}{3}]~$\lambda\lambda$4959,5007)/H$\beta$}
\def\R23{${\rm R}_{23}$}

\newcommand{\Msun}{${\rm M}_{\odot}$}

\newcommand{\NII}{[{\ion{N}{2}}]}

\newcommand{\SIIratio}{[\ion{S}{2}]~$\lambda$6731/[\ion{S}{2}]~$\lambda$6717}

\newcommand{\CIIIf}{[\ion{C}{3}]}
\newcommand{\CIII}{\ion{C}{3}]}

\newcommand{\OH}{$\log({\rm O/H})+12$}

\newcommand{\SII}{[{\ion{S}{2}}]}
\newcommand{\SIII}{[{\ion{S}{3}}]}

\newcommand{\SV}{[{\ion{S}{5}}]}
\newcommand{\NIV}{[{\ion{N}{4}}]}
\newcommand{\NIVi}{{\ion{N}{4}}]}
\newcommand{\NV}{[{\ion{S}{5}}]}
\newcommand{\Hb}{{H$\beta$}}
\def\O4363{[{\ion{O}{3}}]~$\lambda$4363}
\newcommand{\OIII}{[{\ion{O}{3}}]}

\newcommand{\Ha}{{H$\alpha$}}

\def\L60{L$_{60}$}

\newcommand{\rion}[2]{{\ensuremath{\mbox{\rm #1$\,${\sc\expandafter{\romannumeral#2\relax}}}}}}

\shorttitle{ISM pressure and density diagnostics}
\shortauthors{Kewley et~al.}

\begin{document}

\title{Theoretical ISM pressure and electron density diagnostics for local and high-redshift galaxies}

\author{Lisa J. Kewley}
\affil{Australian National University}
\affil{ARC Centre of Excellence for All Sky Astrophysics in 3 Dimensions (ASTRO 3D)}
\email {lisa.kewley@anu.edu.au}

\author{David C. Nicholls}
\affil{Australian National University}
\affil{ARC Centre of Excellence for All Sky Astrophysics in 3 Dimensions (ASTRO 3D)}

\author{Ralph Sutherland}
\affil{Australian National University}
\affil{ARC Centre of Excellence for All Sky Astrophysics in 3 Dimensions (ASTRO 3D)}

\author{Jane R. Rigby}
\affil{NASA Goddard Space Flight Center}

\author{Ayan Acharyya}
\affil{Australian National University}
\affil{ARC Centre of Excellence for All Sky Astrophysics in 3 Dimensions (ASTRO 3D)}

\author{ Michael A. Dopita}
\affil{Australian National University}
\affil{ARC Centre of Excellence for All Sky Astrophysics in 3 Dimensions (ASTRO 3D)}

\author{Matthew B. Bayliss}
\affil{Massachusetts Institute of Technology}

\begin{abstract}
We derive new self-consistent theoretical UV, optical, and IR diagnostics for the ISM pressure and electron density in the ionized nebulae of star-forming galaxies.  Our UV diagnostics utilize the intercombination, forbidden and resonance lines of silicon, carbon, aluminum, neon, and nitrogen.  We also calibrate the optical and IR forbidden lines of oxygen, argon, nitrogen and sulfur.  We show that line ratios used as ISM pressure diagnostics depend on the gas-phase metallicity with a residual dependence on the ionization parameter of the gas.  In addition, the traditional electron density diagnostic \SIIratio\ is strongly dependent on the gas-phase metallicity.  We show how different emission-line ratios are produced in different ionization zones in our theoretical nebulae.  The \SII\ and \OII\ ratios are produced in different zones, and should not be used interchangeably to measure the electron density of the gas unless the electron temperature is known to be constant.  We review the temperature and density distributions observed within \HII\ regions and discuss the implications of these distributions on measuring the electron density of the gas.  Many \HII\ regions contain radial variations in density.  We suggest that the ISM pressure is a more meaningful quantity to measure in \HII\ regions or galaxies.  Specific combinations of line ratios can cover the full range of ISM pressures ($4<\log(P/k)< 9$).  As \HII\ regions become resolved at increasingly high redshift through the next generation telescopes, we anticipate that these diagnostics will be important for understanding the conditions around the young, hot stars from the early universe to the present day.
\vspace{0.5cm}
\end{abstract}

\section{Introduction}
The Interstellar Medium (ISM) pressure and electron density of the gas are two of the main physical quantities that govern the emission from \HII\ regions.  The nebular emission-lines and derived quantities, such as the gas-phase metallicity, ionization parameter, and star formation rate, depend critically on assumptions about the density and pressure of the nebula.   

Direct measurements of the electron density (assuming a constant electron temperature) are now being made for increasing numbers of galaxies, particularly at high redshift.  Some high-z studies find large electron densities \citep{Hainline09,Bian10,Liu08,Brinchmann08b,Shirazi13,Shirazi13b} while others have relatively low electron densities \citep[see][]{Rigby11,Bayliss13}.  Selection effects may play a key role in electron density measurements at high redshift.  \citet{Kaasinen16} showed that the larger electron density seen in many high-z galaxies disappears when high-z and low-z samples are matched in star formation rate.  This result implies that the observed large electron densities are a result of selecting samples of the most luminous galaxies at a given epoch.  

One major disadvantage with measuring the electron density is the assumption of a constant nebular temperature. Under this assumption, the ISM pressure is related to the mean electron temperature through $n =\frac{ P }{ T_e k }$, where the total density $n$ is related to the electron density, $n_e$, through $n=2 n_{e} (1+{\rm He/H})$.  In a fully ionized plasma, the electron temperature is assumed to be $\sim 10^4$~K, and the ISM pressure is then directly proportional to the electron density \citep[see e.g.,][for a discussion]{Dopita06}.   However, in real \HII\ regions, the electron temperature is not constant, and the nebular gas contains temperature fluctuations and gradients, with a complex ionization structure \citep[see e.g.,][for a discussion]{Wang04}.  

Another major disadvantage with measuring the electron density is the assumption that the electron density is a single value across an entire \HII\ region or galaxy.  Both \HII\ regions and galaxies are known to have gradients and clumps in electron density \citep[e.g.,][]{Binette02,Phillips07,Herrera-Camus16}.  In these cases, a constant electron density may not be a realistic assumption.

The ISM pressure provides a useful alternative measure of the ISM conditions.  The ISM pressure is a key determinant of fully self-consistent models that include a detailed nebular temperature and density structure. The pressure is determined by the combination of the mechanical energy produced by the stellar population, as well as the strength and shape of the radiation field.   Models with a constant pressure are applicable when the sound crossing time is less than the heating and cooling timescales, which occurs in the majority of \HII\ regions \citep{Field65,Begelman90}.   

It is now possible to measure the ISM pressure in \HII\ regions and galaxies.  The application of detailed stellar evolution synthesis and photoionization models allows the complex temperature and ionization structure of a nebula to be modeled.  For example,   \citet{Lehnert09} applied photoionization models to galaxies at $z\sim 2$, suggesting that high ISM pressures from large (10-20~kpc) intense star-forming regions exist in most high-z star-forming galaxies. 

In this paper, we calibrate a suite of UV, optical, and IR emission-line ratios to measure the ISM pressure and electron density of the gas.  We use detailed, self-consistent photoionization models that calculate the detailed temperature and ionization structure of nebulae. We investigate the dependence of ISM pressure and electron density diagnostics on the gas-phase metallicity and ionization parameter.  We show that the dependence on the metallicity is large (typically an order of magnitude in $\log(P/k)$), while the effect of ionization parameter is relatively small (typically $<0.1$~dex).  

We provide our model data for calculating the ISM pressure, as a function of metallicity and ionization parameter.  We anticipate that these models and resulting diagnostics will be useful for understanding the physical properties of galaxies across large swathes of cosmic time in the coming era of large-scale infrared spectroscopy with space and 8-10m class ground-based telescopes.

In Section~\ref{models}, we describe the stellar evolution and photoionization models used to calculate the ISM pressure and electron density diagnostics.  We investigate the effect of temperature on the density-sensitive emission-lines in Section~\ref{temperature}.  Section~\ref{pressure} gives the UV, optical, and infrared pressure and electron density diagnostic calibrations for our models.  We explain the dependence of the calibrations on metallicity and ionization parameter as well as the pressure and density regimes over which each diagnostic is valid. In Section~\ref{discussion}, we discuss the ionization and density structure of \HII\ regions and the effect on our diagnostics.  We also briefly discuss the potential contributions to the emission-lines from Diffuse Ionized Gas (DIG) and shocks.  We present our conclusions in Section~\ref{conclusions}.  Throughout this paper, we use Z to denote $\log(O/H)+12$, the gas-phase oxygen abundance\footnote{\citet{Asplund09} gives a bulk solar abundance of \OH$=8.72$, and a surface solar abundance  of \OH$=8.69$.  The solar bulk abundance is an estimate of the abundance when the Sun formed, whereas the solar surface abundance shows evidence of evolution and mixing.  \citet{Nicholls16} argues that local B stars (\OH$= 8.76$) may be more representative of local present-day galactic abundances.}.  We use the term ISM pressure when referring to the pressure within the ionized gaseous nebulae surrounding young hot stars.

\section{Theoretical Models}\label{models}

We combine current stellar evolution synthesis models with our photoionization models to create synthetic spectra of star-forming regions for a broad range of input parameters.  We select our range of parameters to fully cover the expected pressure, density, metallicity and ionization parameter that might exist in galaxies.  

\subsection{Stellar Radiation Field}

We use Starburst99 to generate the stellar ionizing radiation field for input into our photoionization models.
The Starburst99 (SB99) models that we use are described in detail in \citet{Levesque10} and \citet{Nicholls12}.  Briefly, we apply 
a Salpeter IMF \citep{Salpeter55}  with an upper mass limit of 100 \Msun.   The choice of IMF makes negligible difference on the emission-line ratios used in this analysis.  We use the Pauldrach/Hillier model atmospheres, which employ the WMBASIC wind models of \citet{Pauldrach01} for younger ages when O stars dominate the luminosity ($<3$~Myr), and the CMFGEN  atmospheres from \citet{Hillier98} for later ages when Wolf-Rayet (W-R) stars dominate.  These stellar atmosphere models include the effects of metal opacities.   We use the Geneva group ``high" mass-loss evolutionary tracks \citep{Meynet94}.  These tracks include enhanced mass-loss rates that are applicable to low-luminosity W-R stars and can reproduce the blue-to-red supergiant ratios observed in the Magellanic Clouds \citep{Schaller92,Meynet93}.  Starburst99 generates a  synthetic FUV spectrum using isochrone synthesis \citep[e.g.,][]{Charlot91}, in which isochrones are fitted to the evolutionary tracks across different masses rather than discretely assigning stellar mass bins to specific tracks.   To match the nebular metallicities of our photoionization code, we interpolate between the Starburst99 model grids as a function of metallicity.   

\subsection{Photoionization Models}

We use our MAPPINGS version 5.1 photoionization code \citep{Binette85,Sutherland93,Dopita13} to model the interstellar medium surrounding the ionizing radiation field.  

The relative metallicity of individual elements were chosen to maximise compatibility between stellar atmospheres, evolutionary tracks and nebulae. Nebular element abundances at sub-solar metallicities were scaled as described in \citet{Nicholls16}.  Briefly, we use the standard local B-star scale of \citet{Nieva12} for the present day local region of the Milky Way.  This scale is based on an ensemble average of 29 stars, and is most relevant to the present day abundances of the Milky Way local region because B star photospheric abundances represent the bulk abundances of the nebulae in which they recently formed.  The Nieva \& Pryzbilla abundance scale includes the elements He, C, N, O, Ne, Mg, Si, and Fe.  We supplement these elements with recent solar and meteoric abundances \citep{Grevesse15,Scott15a,Scott15b,Lodders09}. Together, these measurements provide local region elemental abundances for 30 elements.  The full abundance sets are given in Tables 1 \& 2 of \citet{Nicholls16}.

The $\alpha$-element abundances are assumed to scale linearly with metallicity, with the exception of helium, carbon, and nitrogen.  The abundance scaling is described in full detail in \citet{Nicholls16}.  For helium, we include the stellar yield in addition to the primordial abundance from \citet{Pagel92}.  For nitrogen, we assume primary and secondary nucleosynthetic component fits to the stellar abundances from \citet{Izotov99}, \citet{Israelian04}, \citet{Spite05}, and \citet{Nieva12}.   For carbon, we assume primary and secondary nucleosynthetic component fits to the stellar abundances from \citet{Gustafsson99}, \citet{Akerman04}, \citet{Spite05}, \citet{Fabbian09}, \citet{Nieva12} and \citet{Nissen14}.

Some metals are depleted out of gas phase and onto dust grains.  For nebular dust depletion we use the parametric models from \citet{Jenkins09}, with a value of Fe depletion of -1.5 dex, derived from Wide Integral Field Spectrograph (WiFeS) observations of local and Magellanic Cloud HII regions (Dopita et al., in prep).   

We use the latest available atomic data for the lightest 30 elements, taken from the CHIANTI 8 database \citep{Delzanna15}.  We use spline interpolation between these values where needed.   
 
MAPPINGS uses either a plane parallel or spherical geometry with the ionization parameter defined at the initial edge of the nebula.  The ionization parameter $q$ has units of velocity (cm/s) and can be thought of as the maximum velocity ionization front that an ionizing radiation field is able to drive through a nebula.  This dimensional ionization parameter is related to the dimensionless ionization parameter $U$ through the identity $U\equiv q/c$.  The dimensionless ionization parameter is $ -3.2 < \log U < -2.9$ for local \HII\ regions \citep{Dopita00} and star-forming galaxies \citep{Moustakas06,Moustakas10}.  

Here, we utilize both spherical and plane parallel models.  Our density and pressure models are calculated using a plane parallel model, to account for the fact that in a real \HII\ region, one typically views an \HII\ region from one direction, not a sum of the light emitted in all directions, i.e. one does not typically view the emission from the far side of the \HII\ region.  We utilize spherical models to investigate how well one can reproduce the density structure of an \HII\ region using our diagnostics.
To minimize small uncertainties produced by particular geometries, we calculate spherical models in which $q$ is determined at the inner radius.   The average ionization parameter is lower than this initial value, and is dependent on the ionization parameter of the initial radius.  In practice, all models with a similar effective ionization parameter produce very similar spectra, assuming all other parameters are held constant \citep[e.g.,][]{Dopita00}.  We have verified that our spherical and plane parallel models used here produce almost identical spectra if they have a similar effective ionization parameter. 

Detailed photoionization, excitation, and recombination are calculated at linear increments (step size $0.02$) throughout the nebula. The model completes when the hydrogen gas is fully recombined.  A full description of the models, including geometry, is given in \citet{Lopez12}.
These models include a sophisticated treatment of dust, including the effects of absorption, grain charging, radiation pressure, and photoelectric heating of the small grains \citep{Groves04a}.  

For the purposes of developing ISM pressure and density diagnostics, we calculate three dimensional grids of models in metallicity, ionization parameter, and a third parameter (pressure, density, or temperature) by holding metallicity fixed at each of \OH$=[7.63, 8.23, 8.53, 8.93, 9.23]$, and the ionization parameter fixed at each of $\log(\frac{q}{cm s^{-1}})=[6.5,6.75,7.0,7.25,7.5,7.75,8.0,8.25,8.5]$, while varying the third parameter.  Using this approach, we develop three sets of three dimensional grids, one set for constant pressures, and one set for constant electron densities, and one set for constant electron temperatures, as described below.

\begin{enumerate}
\item {\it Pressure Models:}  Our pressure models are plane parallel models calculated at constant pressures of $\log(P/k) = 4.0$ to $\log(P/k) = 9.0$ dex in increments of 0.5 dex in $\log(P/k)$, where $P/k$ is in units of $cm^{-3} k$.  This pressure represents the pressure from the mechanical energy flux produced by the stellar population by the combination of stellar winds and supernovae.  The purpose of these models is to calibrate emission-line ratios as a function of ISM pressure.  Therefore, we use the ISM pressure as a variable input parameter into our photoionization models rather than constraining the pressure by the mechanical luminosity of the stellar evolution models, as in \citet{Dopita06}.  Our pressure models calculate a detailed electron temperature and density structure which varies through the nebula based on the metallicity and the ionizing radiation field.  The model metallicities are constrained by the stellar tracks used by the stellar evolutionary synthesis models.  These tracks are based on a coarse model grid with metallicities \OH$=7.63, 8.23, 8.53, 8.93, 9.23$.   The model parameters and resulting fluxes are provided in Table~\ref{pressure_flux_table}.

\item {\it Density Models:}  We calculate constant density plane parallel models with electron density $\log(n_e/{\rm cm^{3}}) = 0$ to $\log(n_e/{\rm cm^{3}}) = 5$ in increments of 0.5 dex, for the model metallicities \OH$=[7.63, 8.23, 8.53, 8.93, 9.23]$.  Like the pressure models, our density models include a changing temperature structure determined by the metallicity and by the ionizing radiation field, but they do not account for a changing density.  Our density models assume a constant density throughout the nebulae.  The model parameters and resulting fluxes from our density models are provided in Table~\ref{density_flux_table}.

\item {\it Temperature Models:}  We calculate single model atoms for the density-sensitive species by fixing the electron density ($\log(n_e/{\rm cm^{3}}) = 0$ to $\log(n_e/{\rm cm^{3}}) = 5$ in increments of 0.5 dex).  For each electron density, we calculate a model for each electron temperature from $\log(T_{e}/K) = 0$ to $\log(T_{e}/K) = 5$ in increments of 0.5 dex.  This approach is different to our Pressure and Density models where temperature structure is an output of the model.  The electron density calibrations in \citet{Osterbrock89} and in the IRAF {\it Temden} task are based on such models, assuming $\log(T_{e}/K) = 4$.  We emphasize that these single atom fixed temperature, fixed density models do not represent the physical conditions in a nebula and they are illustrative rather than diagnostic.  

\end{enumerate}

Of these three sets of models, we consider our Pressure Models to be the most realistic simulations of \HII\ regions because the sound crossing time in a typical \HII\ region nebula is comparable with the cluster lifetime, sufficient for the pressure to equalize across the nebula.  Therefore an \HII\ region can reasonably be assumed to have a constant pressure.  We use our Pressure Models to derive calibrations of ISM pressure for star-forming regions.  We use our Density Models to derive calibrations of the electron density that take the temperature structure of a nebula into account.  The combination of metallicity, ionization parameter, and the shape of the ionizing radiation field produces a complex temperature structure.  Therefore, to isolate the effect of temperature on density-sensitive line ratios, we use our simple temperature models.

\section{The effect of electron temperature}\label{temperature}

To understand the relationship between density-sensitive line ratios and metallicity or ionization parameter, it is important to understand the relationship between emission-line fluxes and the electron temperature within the line-emitting zone for the atomic species being considered.  The electron temperature is sensitive to the metallicity of the gas because metals act as coolants for the nebula.  Therefore, an observed relationship between a line ratio of the same species and metallicity can often be traced to the temperature sensitivity of the emission-lines.  

\subsection{Electron temperature and collisional excitation}

In this paper, we utilize many collisionally-excited lines.  The collisional excitation rate per unit volume ($R_{ij}$ in cm$^{3} {\rm s}^{-1}$) from energy level i to energy level j is given by

\begin{equation}
R_{ij}=n_{e} N_{i} \alpha_{ij},  
\end{equation}

where $n_{e}$ is the electron density per unit volume and $N_{i}$ is the electron density in the lower level.  The collisional coefficient $\alpha_{ij}$ depends on the 
temperature, $T$, of the gas through an exponential thermal excitation term ($e^{\frac{-E_{ij}}{kT}}$) and a power law term ($T^{-1/2}$), i.e.

\begin{equation}
R_{ij}=n_{e} N_{i} \left( \frac{2 \pi \hbar^4}{k m_{e}^3}  \right)^{1/2} \frac{\Omega_{ij}}{g_{i}}\,  T^{-1/2} exp \left( \frac{-E_{ij}}{kT} \right) \label{eq_R12}
\end{equation}

where $\Omega_{ij}$ is the collision strength for the transition from level i to level j, $g_{i}$ is the statistical weight of level i, and $E_{ij}$ is the energy difference between levels i and j.  The relative importance of each of the two temperature terms governs whether a line ratio pair is sensitive to the electron density of the gas.

The upper energy level can be depopulated by collisional de-excitation or by radiation.  The radiative depopulation rate is given by

\begin{equation}
R_{ji}^{rad}=n_{e} N_{j} A_{ji}
\end{equation}

where $A_{ji}$ is the spontaneous transition probability and $N_{j}$ is the number of electrons in the upper energy level.

The collisional de-excitation rate is given by

\begin{equation}
R_{ji}^{coll}=n_{e} N_{j} \left( \frac{2 \pi \hbar^4}{k m_{e}^3}  \right)^{1/2} \frac{\Omega_{ji}}{g_{j}}\,  T^{-1/2} \label{eq_R21}
\end{equation}

where $\Omega_{ji}$ is the collision strength for the transition from level j to level i, and $g_{j}$ is the statistical weight of level j.  The collision strength $\Omega_{ji}$ has a residual dependence on temperature that becomes important for the mid-IR fine structure lines.  

The critical density is a useful parameter when considering the emission-line strengths from \HII\ regions.  The critical density is defined as the density where the collisional de-excitation probability equals the radiative de-excitation probability for the excited state.

\subsection{Density-sensitive line ratios and electron temperature}\label{density_temperature}

The most commonly used density diagnostics are the optical line ratios of [\rion{S}{2}] and [\rion{O}{2}]. However, there are many potential choices for emission line pairs that can be density diagnostics. We use our simple temperature models as a guide to determine (a) which ratios may be used as density diagnostics, (b) the range of electron densities over which the ratios are useful, and (c) their dependence on the electron temperature.   Table~\ref{table_ne} gives the line ratios considered here, their ionization potentials, and the range over which each ratio can be used as a reliable density or ISM pressure diagnostic.

\begin{figure*}[!t]
\epsscale{1.0}
\plotone{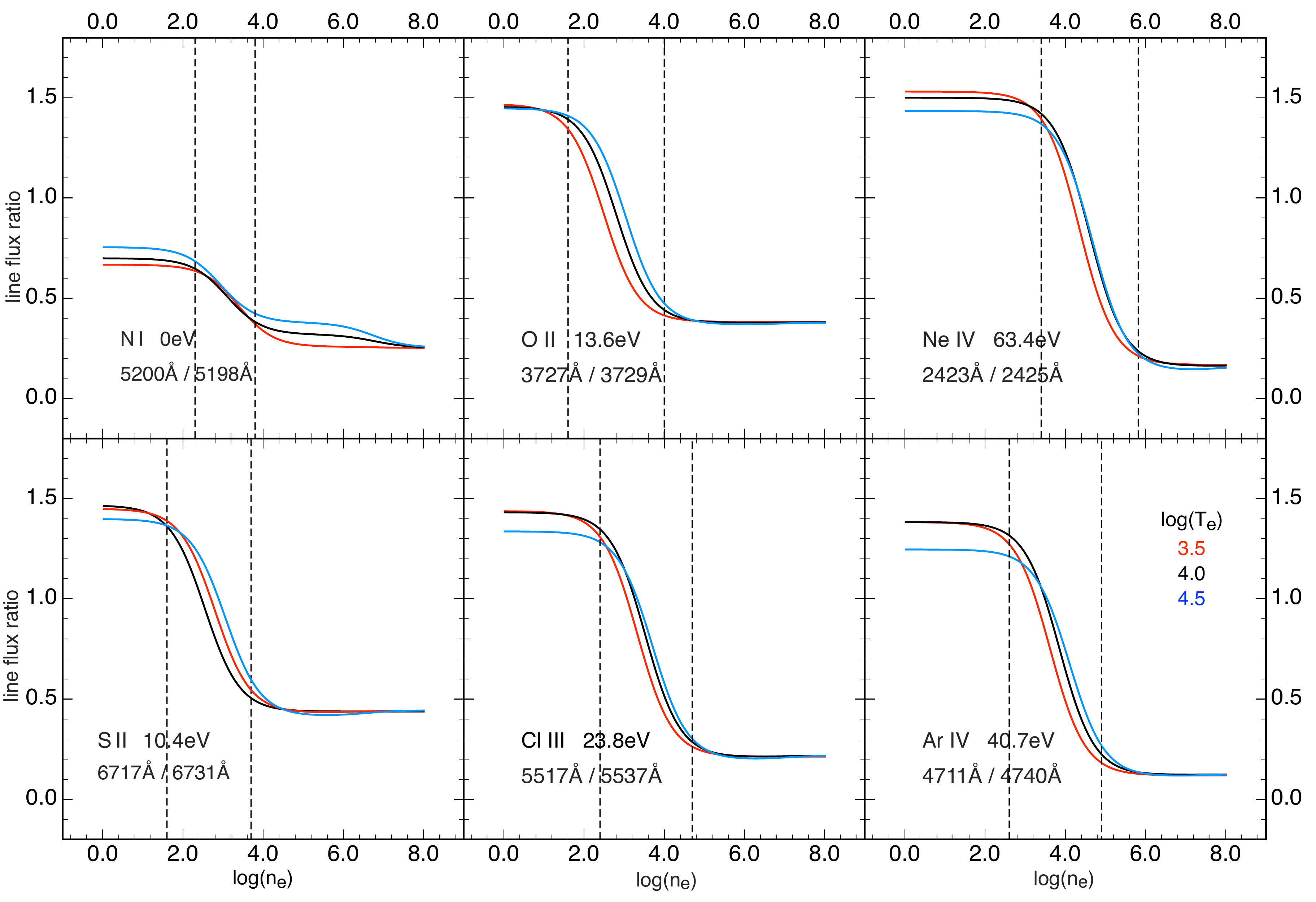}
\caption[fig1.pdf]{Emission line flux ratios versus electron density for Case 1 line ratios \rion{N}{1}, \rion{O}{2}, \rion{Ne}{4}, \rion{S}{2}, and \rion{Cl}{3}, and \rion{Ar}{4}.  Case 1 ratios are based on forbidden lines that have low radiative transition probabilities.  The curves were were calculated with our MAPPINGS v5.1 code using single atom data for electron densities between 1 and 10$^8$ cm$^{-3}$, and electron temperatures for log(T$_e$) = 3.5 (red), 4.0 (black) and 4.5 (blue), covering the conditions likely to be encountered in \rion{H}{2} regions.  The vertical dashed lines mark the range of electron densities for which the ratio is a useful density diagnostic.}
\label{scenario1_ratios}
\end{figure*}

There are three different scenarios where useful nebula density diagnostic line pair ratios might be expected:

{\bf Case 1:}  The flux ratio of two forbidden lines of the same atom where both lines have low radiative transition probabilities, $A_{ji}$ (s$^{-1}$), giving rise to critical densities that are also low.   Here, the radiative transition probability $A_{ji}$ is small, and is comparable to the collisional de-excitation rate at low densities (where forbidden transitions occur).  In this case, the line fluxes depend on $n_{e}^2$ at low density and on $n_{e}$ at high density, giving a double exponential shaped curve for the line ratios between these two limits.  In this scenario, the two atomic transitions arise from closely spaced upper energy levels, where the energy difference between these closely spaced levels is small compared to their energies above the ground state.  The thermal excitation term $\exp(E_{ij}/kT)$ (i.e. the exponential term in equation~\ref{eq_R12}) is similar for the two lines.  Therefore, the two energy levels are populated by a similar number of electrons at a given temperature, reducing the temperature sensitivity of the line ratio.  In this case, the de-excitation rate is only dependent on $T^{-1/2}$ from the Maxwell-Boltzmann electron velocity distribution, and the radiative depopulation rate is constant ($A_{ji}$ s$^{-1}$).  

Figure~\ref{scenario1_ratios} gives examples of how Case 1 line ratios depend on electron density and temperature, using our Temperature Models.  All line ratios show the typical double exponential shaped curve, with a strong dependence on electron density.  The line ratios typically vary by 1 to 1.5~dex with electron density, and can probe different density regimes.  Each line ratio has a unique dependence on temperature (shown as coloured curves in Figure~\ref{scenario1_ratios}).  Multi-level effects cause the low density limit to be sensitive to the electron temperature for the majority of these line ratios (with the exception of the \OII\ ratio).    Varying the temperature from $\log({\rm T_{e}}) = 3.5$ to $\log({\rm T_{e}}) = 4.5$ causes a small change ($<0.15$~dex) change in the line ratios between the low and high density limits.

The high excitation species such as \ArIV\ and \NeIV\ are particularly sensitive to the electron temperature, and these lines typically arise from high excitation sources such as WR stars or AGN.  The presence of an AGN would need to be ruled out prior to the use of these lines as pressure or density diagnostics because our diagnostics assume a stellar ionizing radiation field only.

The \rion{N}{1} ratio shows a different dependence on electron density and temperature to the other Case 1 line ratios, and is only a weak function of electron density.  Unfortunately, the \NI\ ratio suffers from additional collisions between the two upper energy levels, as well as transitions from higher energy levels, reducing the sensitivity of this line ratio to the electron density.

{\bf Case 2:}  The flux ratio of one line with a small radiative transition probability $A_{ji}$ and one line with a large radiative transition probability from the same atom.  
The line with a small radiative transition probability will be weak, with a low critical density, and will be predominantly collisionally de-excited.  For this line, the line flux will be proportional to $n_{e}$.  The line with a large radiative transition probability will be strong, with a high critical density.  For ISM densities less than this high critical density, this stronger line will be produced predominantly by radiative de-excitation, and the line flux will be proportional to $n_{e}^2$.  The ratio of these two lines produces a low density limit, but the flux ratio at the high density limit is extremely sensitive to the electron temperature.  The ratio of forbidden lines to intercombination lines show this type of behaviour.  These ratios can be used to measure electron density wherever the forbidden line is detectable. 

More accurate densities will be achieved when the electron temperature has been constrained prior to or in conjunction with the application of these density diagnostics.  In Nicholls et al. (in prep), we present electron temperature diagnostics for temperature-sensitive line ratios in the UV.  We note that the electron temperature is different for different regions of the nebula.  Therefore, the electron temperature should be estimated in the zone where the density-sensitive lines are produced.  

\begin{figure*}[!t]
\epsscale{1.0}
\plotone{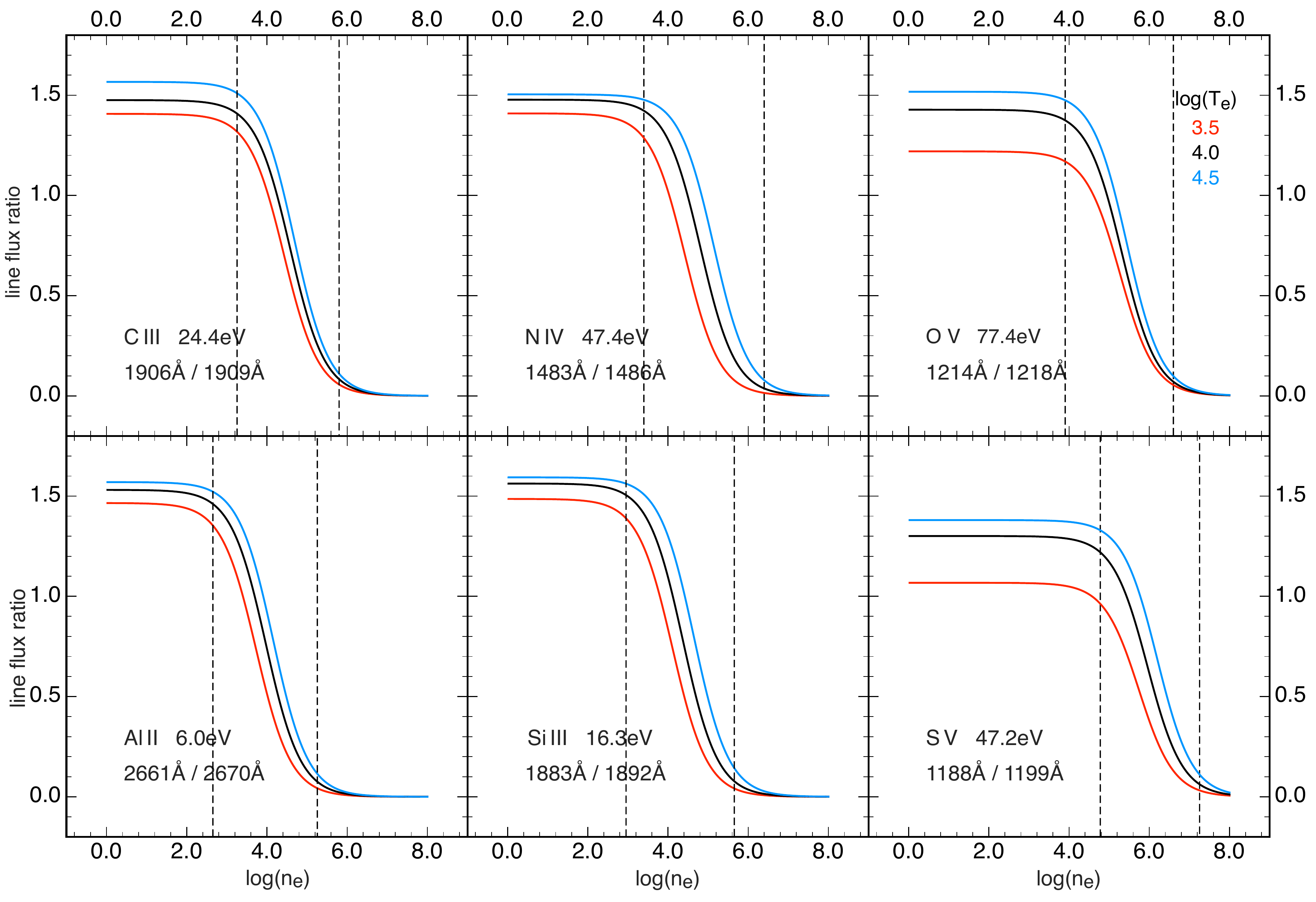}
\caption[fig2.pdf]{Emission line flux ratios for  \rion{C}{3}, \rion{N}{4}, \rion{O}{5}, \rion{Al}{2}, \rion{Si}{3} and \rion{S}{5}, calculated with MAPPINGS v5.1 using single atom data for electron densities between 1 and 10$^8$ cm$^{-3}$, and electron temperatures for log(T$_e$) = 3.5, 4.0 and 4.5, covering the conditions likely to be encountered in an \rion{H}{2} region.  Dashed lines show the range over which the line ratios provide reliable estimates of the electron density.  The density diagnostics for Case 2 are sensitive to the electron temperature within these ranges and should ideally be applied alongside an electron temperature diagnostic.}
\label{scenario2_ratios}
\end{figure*}

In Figure~\ref{scenario2_ratios}, we give examples of emission-line ratios that are produced via Scenario 2.  The low excitation species \AlII, \SiIII, and \CIII\ are produced at densities of typical \HII\ regions, while the higher excitation species (\NIV\ and \OV) can be produced in the ISM surrounding WR stars or an AGN.  

\begin{figure*}[!t]
\epsscale{1.0}
\plotone{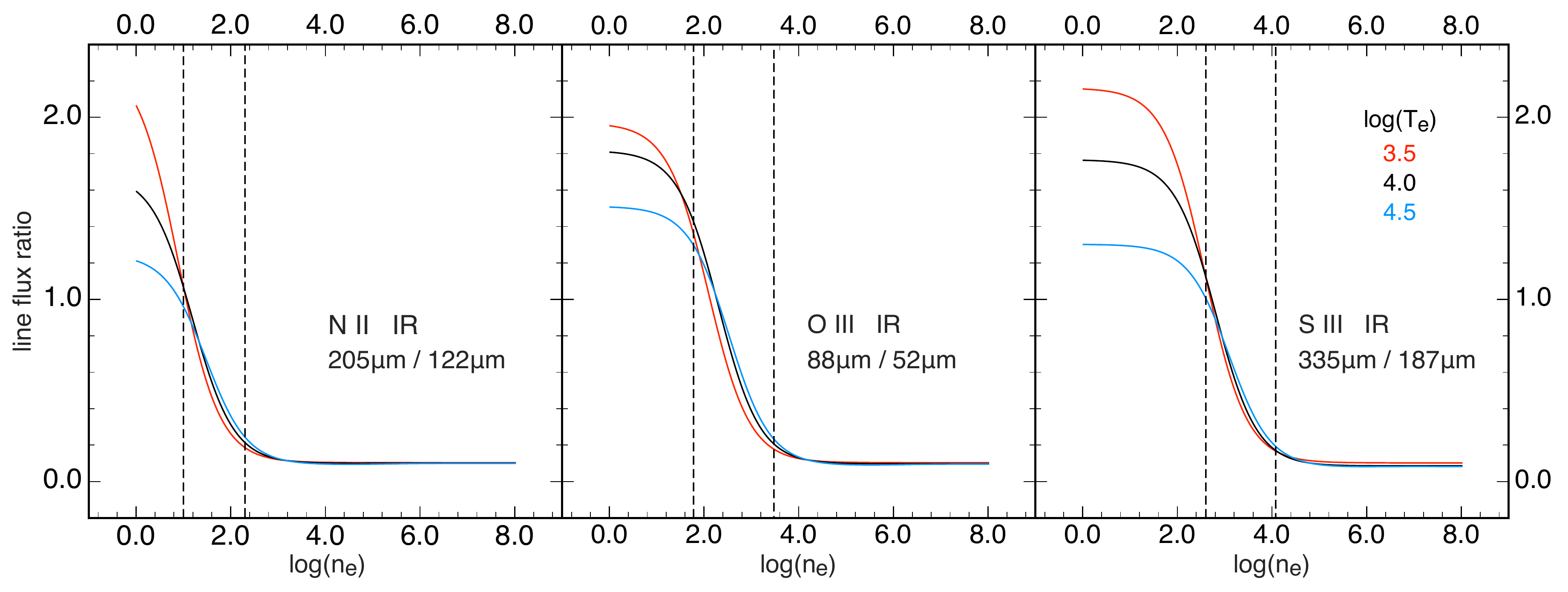}
\caption[fig3.pdf]{Emission line flux ratios for  \rion{N}{2}, \rion{O}{3} and \rion{S}{3}, calculated with MAPPINGS v5.1 using single atom data for electron densities between 1 and 10$^8$ cm$^{-3}$, and electron temperatures for log(T$_e$) = 3.5, 4.0 and 4.5, covering the conditions likely to be encountered in an \rion{H}{2} region.  These density diagnostics for Case 3 line ratios are insensitive to electron temperature for the range over which the line ratios provide reliable estimates of the electron density (dashed lines).  The low density limit depends strongly on electron temperature.}
\label{scenario3_ratios}
\end{figure*}

{\bf Case 3:} The flux ratio of two low energy infrared lines for which the temperature term $\exp(-E_{ij}$/kT) $\sim$ 1 (i.e. $E_{ij} << kT$).  Both collisional excitation and de-excitation are then dominated by the $T^{-1/2}$ term in equations~\ref{eq_R12} and \ref{eq_R21}.  This gives only a very weak dependence on temperature except in the low density limit, and therefore these line ratios are ideal tracers of the electron density of the gas.  The triplet fine structure levels in many ions satisfy this criteria, with low critical densities similar to densities in \rion{H}{2} regions.  The disadvantage of these line ratios is that the energy levels differ significantly, producing lines that are separated in wavelength by a few tens to hundreds of microns.   IR satellites with a large wavelength coverage are often required to reliably measure these line ratios to avoid errors from aperture effects and relative  flux calibration issues.  In Figure \ref{scenario3_ratios}, we give the relationship between the IR line ratios [\rion{N}{2}], [\rion{O}{3}] and [\rion{S}{3}] and electron density, as a function of temperature.    In the low density limit, the line ratios are an extremely strong function of temperature.   The dashed lines in Figure~\ref{scenario3_ratios} indicate the regime over which the line ratios can be used to reliably estimate the electron density.   The IR line ratios each cover a relatively small, but complementary range of electron densities.  Together, these three line ratios can probe the range of electron densities seen in most \HII\ regions and galaxies.

\section{Measuring the ISM Pressure and Density}\label{pressure}

Recent integral field spectroscopic studies of local \HII\ regions show that the gas within \HII\ regions spans a range of temperatures, with electron temperature gradients (Zahid et al. 2017, in prep) and/or fluctuations \citep[see][for a review]{Stasinska02}.  Assuming a constant temperature becomes particularly problematic when fixed-size apertures capture the light from an ensemble of \HII\ regions.  Nuclear or global fiber-based or slit-based spectra of nearby or distant galaxies represent the luminosity-weighted average of multiple (up to hundreds or thousands) of \HII\ regions, each of which may contain a complex electron temperature structure.  In these cases, it is unclear what an electron density that assumes a fixed electron temperature means for the ensemble.  

Electron temperature gradients also exist in spiral galaxies \citep{Quireza06}.  The electron temperature is extremely sensitive to the metallicity of the gas.  Disk galaxies often have steep metallicity gradients that cause a radial variation in the electron temperature \citep[see e.g.,][for a review]{Shields90}.  However, not all galaxies have strong metallicity gradients.  Many barred spiral galaxies have weaker metallicity gradients than spirals of similar type \citep[e.g.,][]{Pagel79,Roy97}, and mergers and irregular galaxies have flat gradients compared with isolated spiral galaxies \citep[e.g.,][]{Kewley10,Rupke10b,Rich12}.  These results suggest that radial gas flows suppress or mix metallicity gradients \citep[e.g.,][]{Roberts79,Martin94,Roy97,Rupke10,Torrey12}.  A constant temperature might be a reasonable assumption in some of these cases.

\subsection{Theoretical density and temperature distributions}

The optical density-sensitive line ratios have traditionally been calibrated assuming a simple model atom and a single electron temperature (typically $T_e = 10^4$~K) \citep[e.g.,][]{deRobertis87,Osterbrock89,Rubin94,Shaw95}.  More recently, calibrations have been based on photoionization models, which assume a constant temperature throughout the photoionized nebula \citep[e.g.,][and references therein]{Proxauf14}.  Our models take into account all available atomic levels, rather than simple model level atoms, and our pressure models allow the electron temperature and electron density to vary through the nebula. 

\begin{figure*}[!t]
\epsscale{1.2}
\plotone{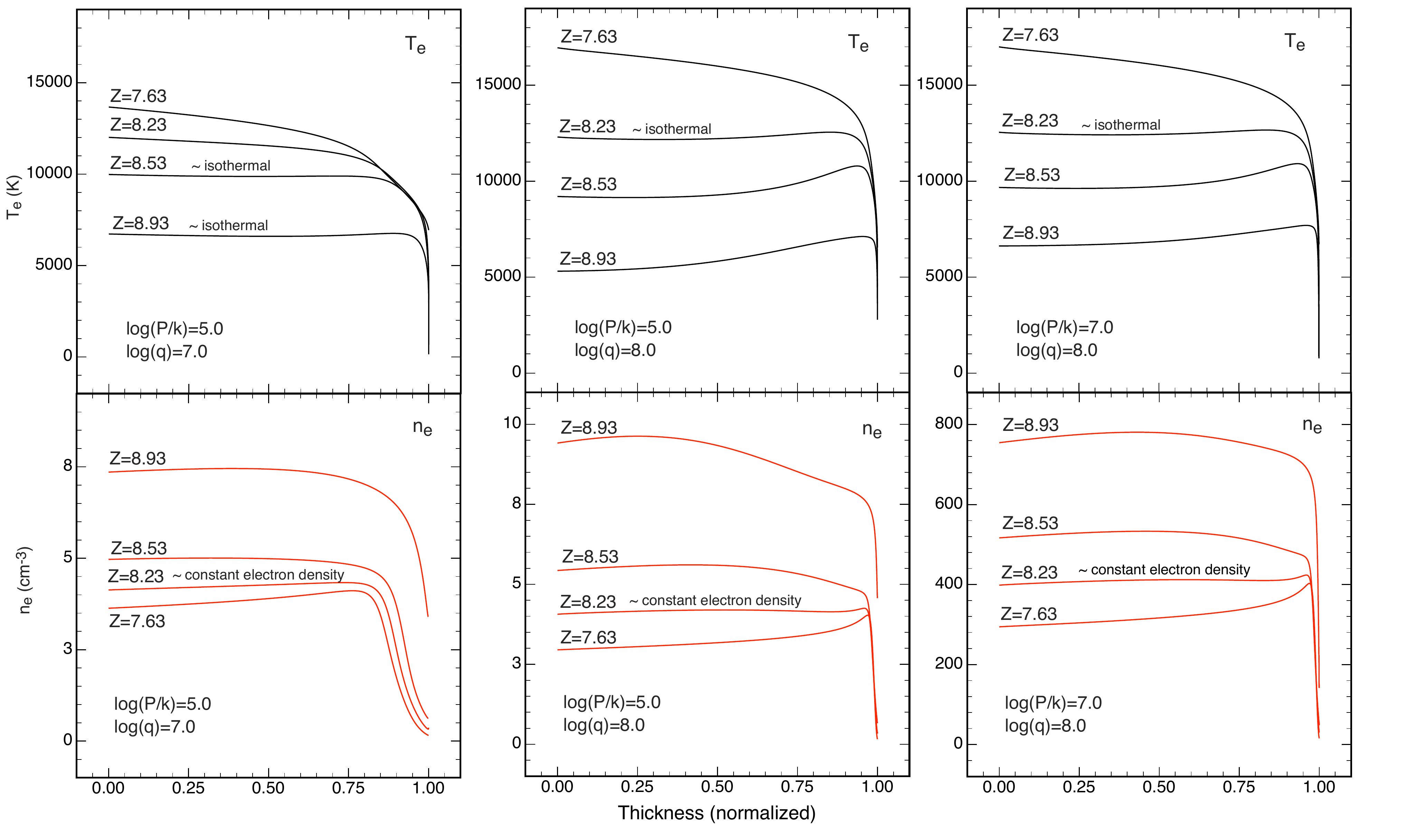}
\caption[fig4.pdf]{Electron temperature ($T_e$) and electron density ($n_e$) as a function of shell thickness for our plane parallel constant pressure models.   Columns show different pressure and ionization models: $\log(P/k)=5.0$, $\log(q)=7.0$ (left),  $\log(P/k)=5.0$, $\log(q)=8.0$ (middle), and $\log(P/k)=7.0$, $\log(q)=8.0$ (right).  
}
\label{Temp_density_structure}
\end{figure*}

Figure~\ref{Temp_density_structure} shows how the electron temperature and density vary through the nebula for our pressure models for a variety of pressure, ionization parameters (different panels), and metallicity (coloured curves).  The electron temperature and density vary with distance through the nebula, and only reflect isothermal or constant electron density conditions for some metallicities (usually, but not always, \OH$\sim 8.23$).  At metallicities at or above \OH$\sim 8.5$, the nebula becomes hotter towards the outer edges.  The cause of this heating is twofold; (1) in the outer zones, the soft ionizing photons have already been absorbed by metals closer to the ionising source, leaving predominantly hard ionising photons, yielding more heat per ionisation, and (2) the dominant coolant, \OIII, dominates the cooling in the inner nebula zone, but is not a significant coolant in the outer edge of the nebula.  A metal-rich nebula absorbs more soft ionizing photons, as well as cooling more strongly in the inner nebula, causing a larger temperature difference between the inner zone and the outer edge of the nebula.

\subsection{UV Pressure and Density Diagnostics}

The UV contains relatively few ISM pressure diagnostics.  Some pressure-sensitive lines (such as the \CII\ complex at 2326\AA\ and the \NIII\ complex at $1746-1753$\AA) are strongly blended and do not have useful diagnostic capability with modern instruments.  Three of the five \CII\ lines between 2323\AA\ and 2329\AA\ are separated by 1\AA\ or less.  The \CII\ lines may also be blended with \ion{Fe}{2} at 2327\AA\ and \ion{Si}{2} at 2329\AA, depending on the metallicity of the gas.  Similarly, three of the five \NIII\ lines between 1746\AA\ and 1756\AA\ are also separated by 1\AA\ or less. 

The \OV\ and \SV\ emission-lines shown in Figure~\ref{scenario2_ratios} are not produced in our pressure models for star-forming regions at detectable levels (i.e. with fluxes $>1\times 10^{-5}$ \Hb).  If the \OV\ and \SV\ lines are observed in a spectrum, an AGN or shocks are likely to be present in the galaxy.  In this Section, we provide UV pressure and density diagnostics for sets of emission-lines that are currently available, or are likely to be available in the near future.

\subsection{The \SiIIIf,\SiIII\ ratio}

 The \SiIIIf~$\lambda 1883$ and \SiIII~$\lambda 1892$ lines are produced by the ${}^{1} S_{0} \rightarrow {}^{3}P_{2}$ and  ${}^{1}S_{0} \rightarrow {}^3 P_{1}$ transitions, respectively. 
The \SiIIIf~$\lambda 1883$/\SiIII~$\lambda 1892$ ratio probes high density, high pressure regions of the nebula ($\log({\rm P/k}) > 7.5$).  \citet{Dufton84} first proposed these \ion{Si}{3} lines as density diagnostics.  This ratio has been used to derive electron densities in planetary nebulae \citep{Dufton84,Clegg87}, and more recently to derive the electron density in the gas around quasars \citep{Negrete12}.  

\begin{figure*}[!t]
\epsscale{1.1}
\plotone{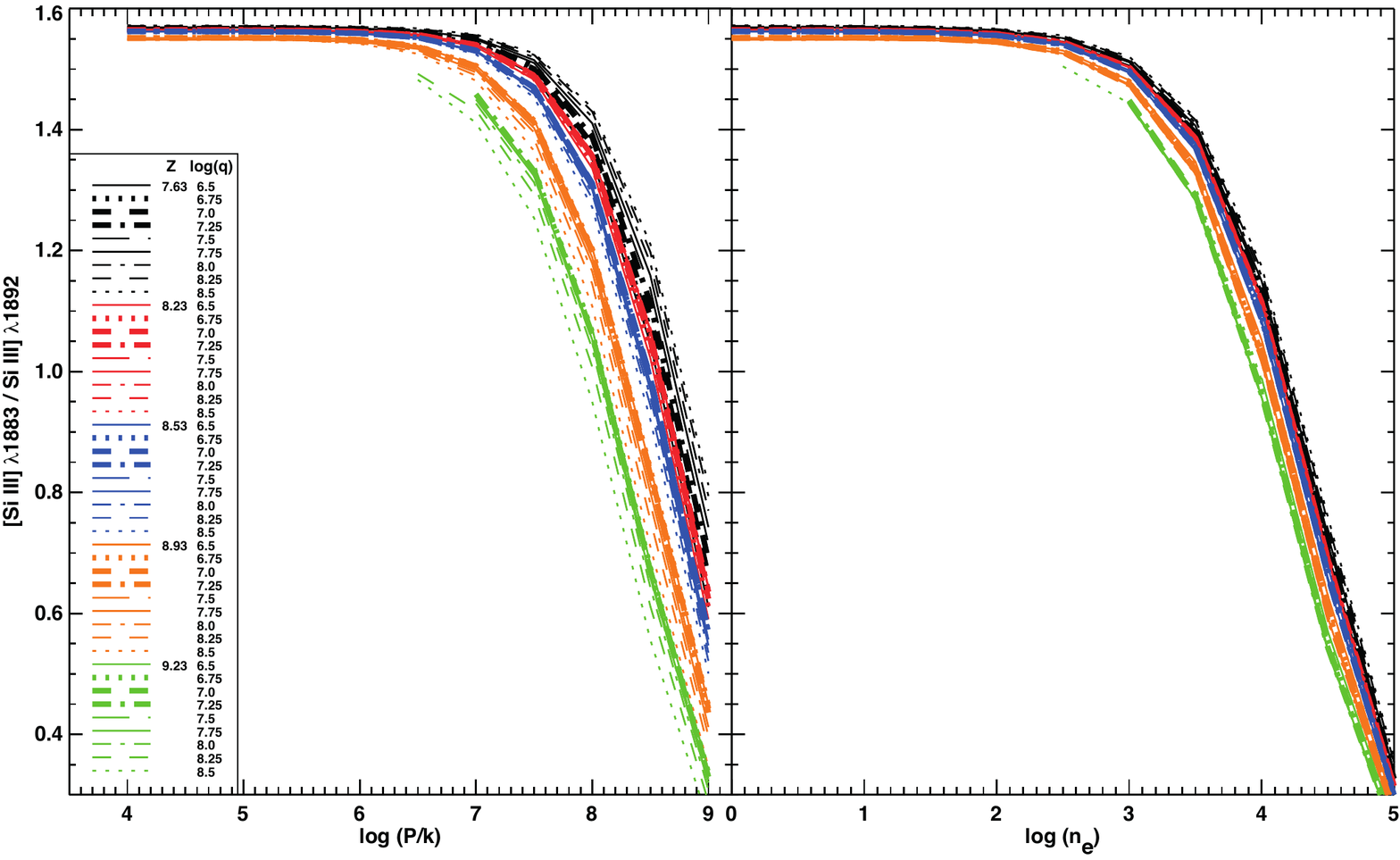}
\caption[fig5.eps]{The theoretical relationship between the \SiIIIf~$\lambda 1883$/\SiIII~$\lambda 1892$ ratio and the ISM pressure (left) and electron density (right) for the metallicities (colored curves) and ionization parameter (solid, dotted, dashed, and dot-dashed lines) covered by our models, as shown in the legend.  The relationship between the \SiIII\ line ratio and the ISM pressure depends primarily on the metallicity, with a small effect from the ionization parameter.  The metallicity (Z) shown in the legend is in units of \OH.}
\label{SiIII_pressure}
\end{figure*}

Figure~\ref{SiIII_pressure} indicates how the \SiIIIf/\SiIII\ ratio depends on ISM pressure, metallicity, and ionization parameter.  The \SiIIIf/\SiIII\ ratio 
is extremely sensitive to the ISM pressure in the high pressure regime, varying by over an order of magnitude between $7.5 <\log({\rm P/k}) < 9.0$.  For the same \SiIIIf/\SiIII\ ratio, the estimated ISM pressure can vary by up to $\sim 1$~dex when the metallicity is changed from \OH$=7.63$ to \OH$=9.23$.  
The estimated pressure also has a minor dependence on ionization parameter ($\sim 0.2$~dex).   Both of our ISM pressure and density calibrations are only valid for  \SiIIIf/\SiIII\ ratios below \SiIIIf/\SiIII$<1.4$, above which the \SiIIIf/\SiIII\ ratio saturates (called the low density limit).

\subsection{The \CIIIf,\CIII\ ratio}

The \CIIIf~$\lambda 1907$/\CIII~$\lambda 1909$ ratio was first proposed as a density diagnostic by \citet{Nussbaumer79}.  Revised density calibrations were calculated by \citet{Keenan92}.  The \CIIIf~$\lambda 1907$ and \CIII~$\lambda 1909$ lines are based on the $ {}^{1} S_{0} \rightarrow {}^{3}P^{o}_{1}$ and the $ ^{1} S_{0} \rightarrow {}^{3}P^{o}_{2}$ transitions, respectively.  The \CIII~$\lambda 1909$ line is an intercombination line, while the weaker \CIIIf~$\lambda 1907$ line is forbidden.  There is a large difference in lifetimes for these transitions, producing a large difference in critical densities and therefore the ratio is extremely sensitive to the electron density of the gas.

\begin{figure*}[!t]
\epsscale{1.0}
\plotone{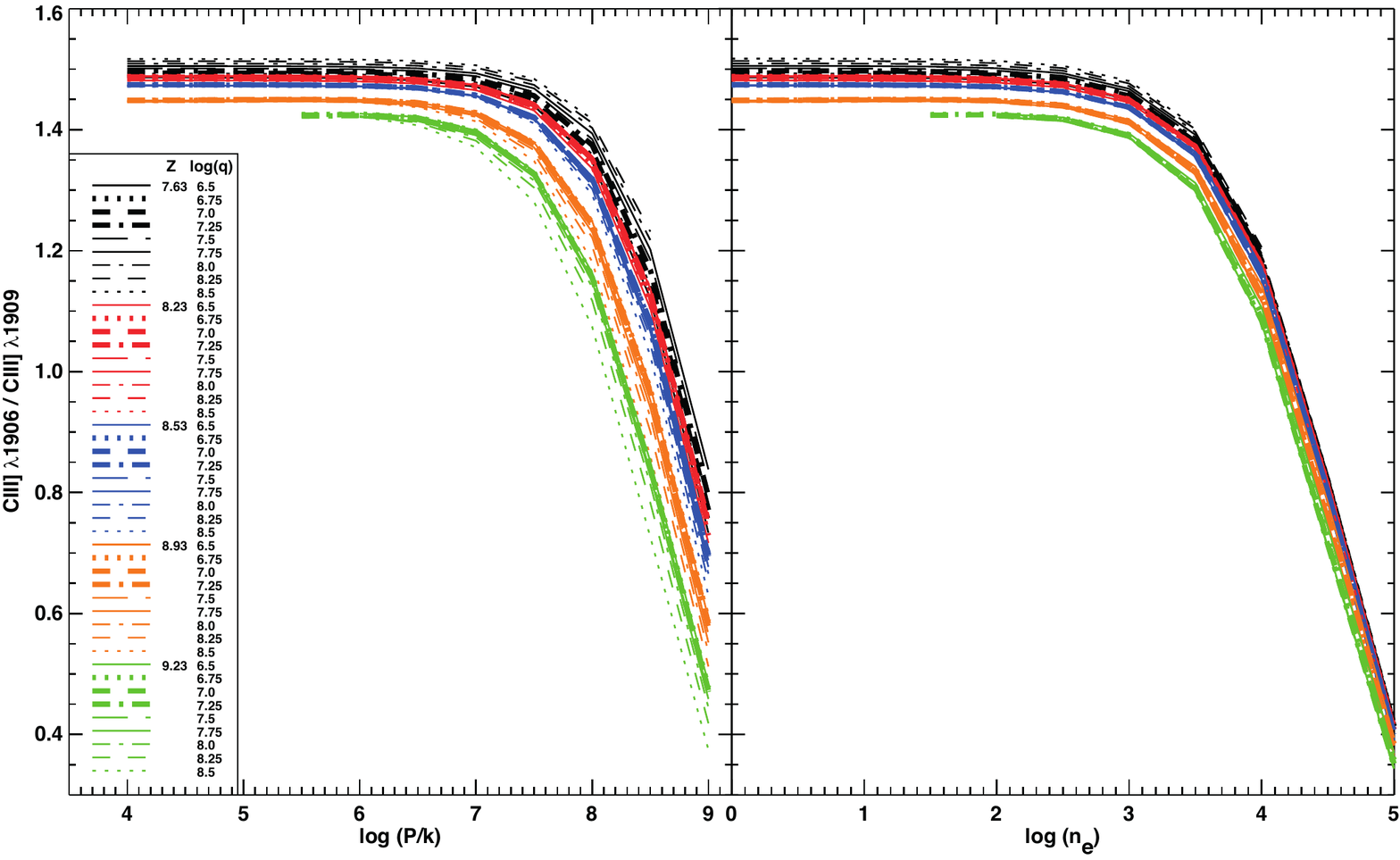}
\caption[fig6.eps]{The theoretical relationship between the \CIIIf~$\lambda 1906$/\CIII~$\lambda 1909$ ratio and the ISM pressure (left) and the electron density (right) for the metallicities (colored curves) and ionization parameter (solid, dotted, dashed, and dot-dashed lines) covered by our models, as shown in the legend.  The relationship between the \CIII\ line ratio and the ISM pressure depends primarily on the metallicity, with a small effect from the ionization parameter.  The metallicity (Z) shown in the legend is in units of \OH.}
\label{CIII_pressure}
\end{figure*}

Figure~\ref{CIII_pressure} shows how the \CIIIf/\CIII\ ratio depends on ISM pressure (left panel) and electron density (right panel) as a function of metallicity and ionization parameter.  The \CIIIf/\CIII\ lines probe the high density, high pressure regions of a nebula ($\log({\rm P/k}) > 7.5$).  In this high pressure regime, the \CIIIf/\CIII\ ratio is an extremely sensitive function of pressure or density, with the \CIIIf/\CIII\ ratio changing by almost an order of magnitude between $7.5 <\log({\rm P/k}) < 9.0$.  
At lower pressures ($\log({\rm P/k}) < 7.5$), the \CIIIf/\CIII\ ratio is primarily sensitive to the metallicity.  Like \SiIII, for the same \CIIIf/\CIII\  ratio, the estimated ISM pressure can vary by up to $\sim 1$~dex when the metallicity is changed from \OH$=7.63$ to \OH$=9.23$.  
The estimated pressure also has a minor dependence on ionization parameter ($\sim 0.2$~dex).   The metallicity dependence is primarily caused by the fact that the C$^{++}$ transitions depend on the electron temperature of the gas through their recombination coefficients.  

The \CIIIf/\CIII\ ratio is a sensitive electron density diagnostic for dense environments ($\log( n_{e}) > 3.5$), where the electron density is uniform throughout the nebula.  Metallicity variations may introduce a scatter in the estimated density of up to 0.3 dex.  For \CIIIf/\CIII\ ratios $>1.4$, the \CIIIf/\CIII\ ratio is in its low density limit, where every collisional excitation is followed by the emission of a photon.  In this regime, the \CIIIf/\CIII\ ratio is not a useful pressure or density indicator.

\subsection{The \AlII\ , \AlIIint\ ratio}

The \AlII~$\lambda 2660$ and \AlIIint~$\lambda 2669$ lines are produced by the ${}^3 P^0 _2   \rightarrow  {}^1 S_0$ and ${}^3 P^{0}_{0} \rightarrow {}^1 S_{0}$ transitions.  The \AlIIint~$\lambda 2669$ intercombination line was first predicted by \citet{Aller61}, and subsequently observed in planetary nebulae \citep{Gurzadian75}.  The \AlII~$\lambda 2660$ transition produces a forbidden line, which has been observed in nearby \HII\ regions \citep{Park10}.

\begin{figure*}[!t]
\epsscale{1.0}
\plotone{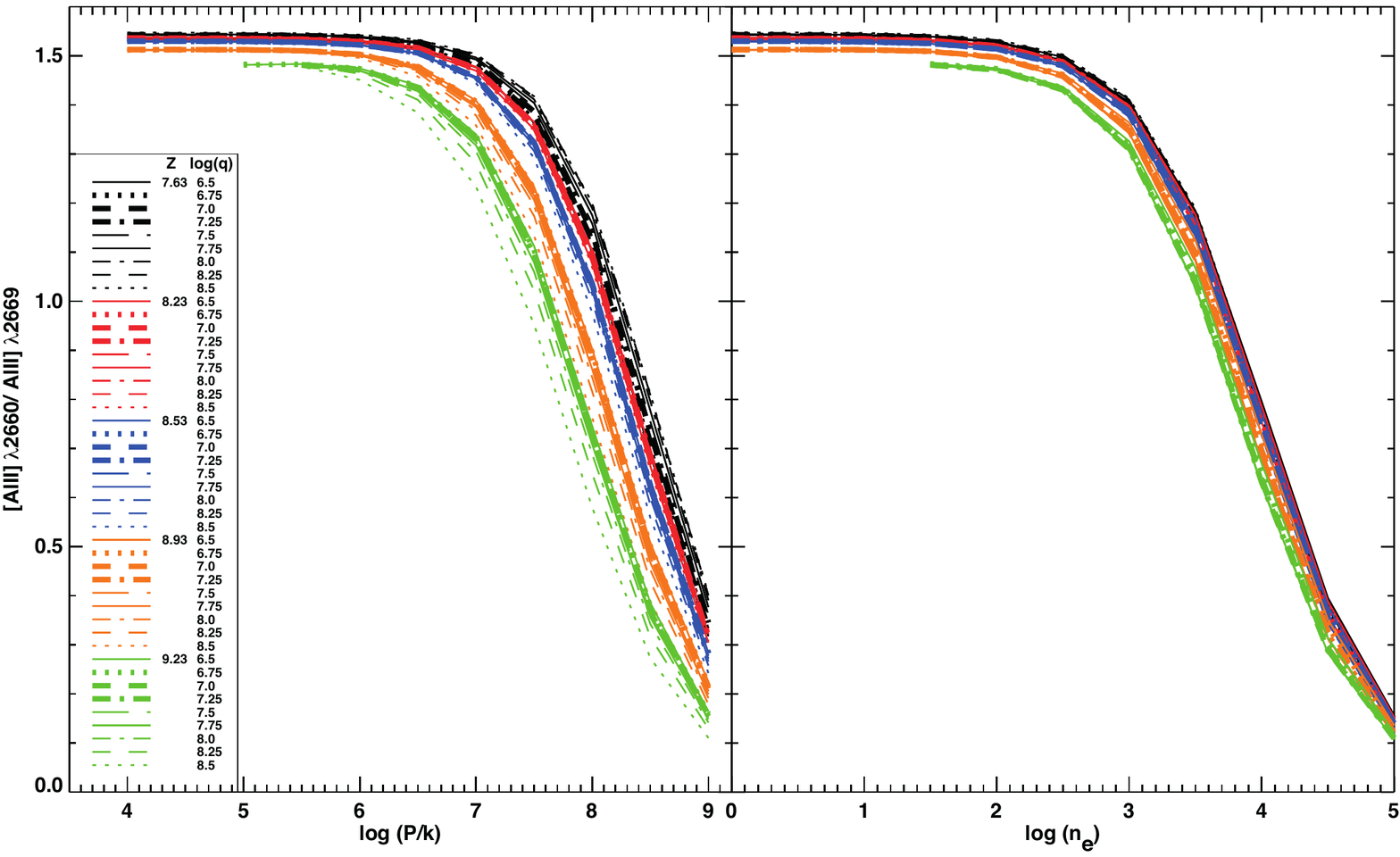}
\caption[fig7.eps]{The theoretical relationship between the  \AlII~$\lambda 2660$/\AlIIint~$\lambda 2669$ ratio and the ISM pressure (left) and electron density (right) for the metallicities (colored curves) and ionization parameter (solid, dotted, dashed, and dot-dashed lines) covered by our models, as shown in the legend.  The relationship between the \AlII\ line ratio and the ISM pressure depends primarily on the metallicity, with a small effect from the ionization parameter.  The metallicity (Z) shown in the legend is in units of \OH.}
\label{AlII_pressure}
\end{figure*}

In Figure~\ref{AlII_pressure} we show that the relationship between the \AlIIres\ ratio and pressure is a sensitive function of metallicity, with a secondary dependence on ionization parameter ($<0.1$~dex).  The \AlIIres\ ratio is useful in probing the high pressure regime, varying by over an order of magnitude between $7.0 <\log({\rm P/k}) < 9.0$.   Both of our ISM pressure and density calibrations are only valid for  \AlIIres\ ratios below \AlII$/$\AlIIint$<1.44$, above which the \AlIIres\ ratio is in its low density limit.  The value at which the \AlIIres\ ratio enters the low density limit depends on metallicity, as a result of the electron temperature dependence for Case 2 flux ratios.

\subsection{The high-ionization \NeIV\ and \NIV,\NIVi\ ratios}

Two high ionization UV line ratios are produced by our stellar photoionization models. These ratios probe the highest pressure regions of a nebula ($\log(P/k) > 8.0$~dex), providing a set of complementary diagnostics to the \CIII\ and \AlII\ lines, which probe lower pressure regions of a nebula.

\begin{figure*}[!t]
\epsscale{1.0}
\plotone{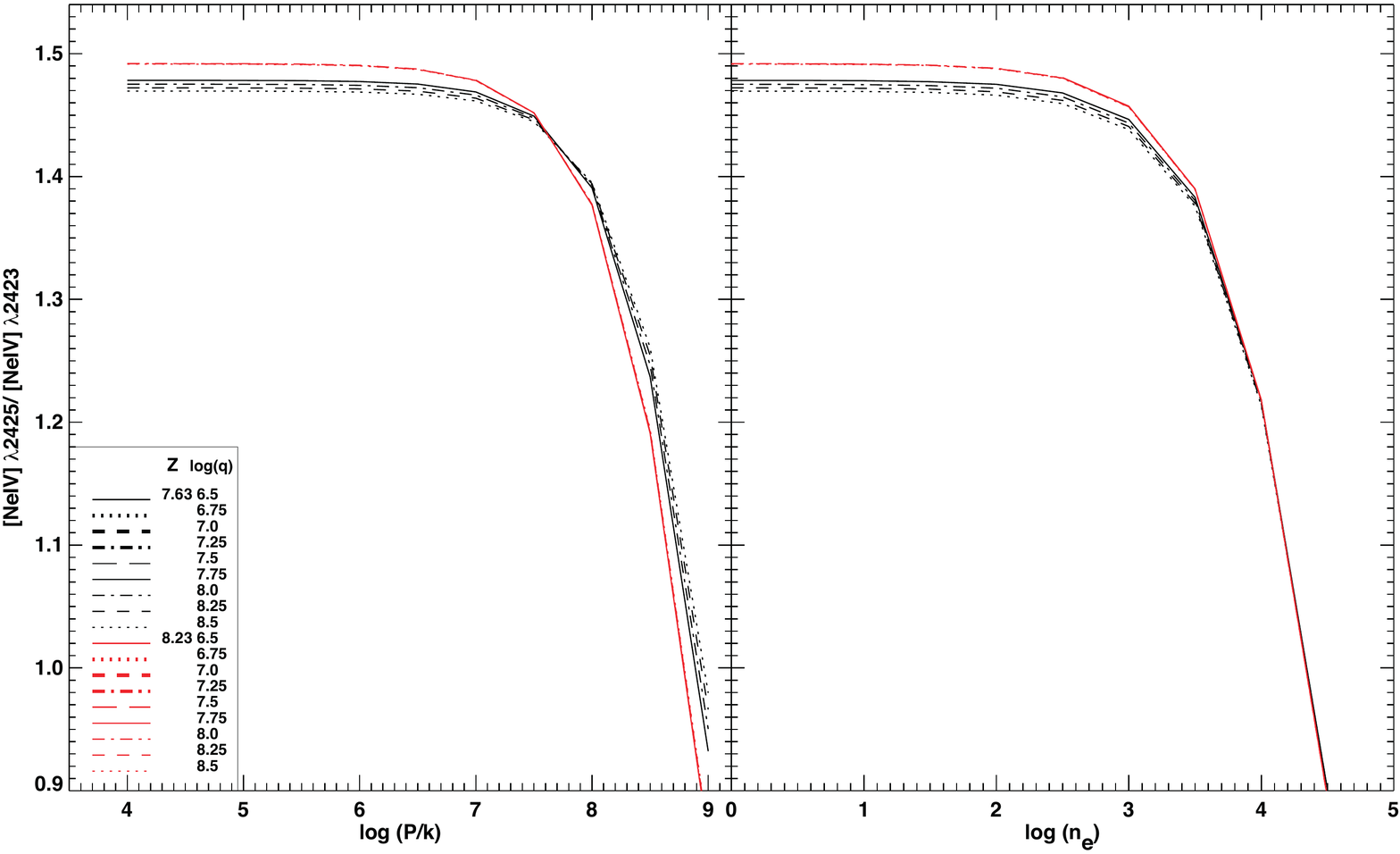}
\caption[fig8.eps]{The theoretical relationship between the \NeIV~$\lambda 2425$/\NeIV~$\lambda 2423$ ratio and the ISM pressure (left) and electron density (right) for the metallicities (colored curves) and ionization parameter (solid, dotted, dashed, and dot-dashed lines) covered by our models, as shown in the legend where Z$=\log({\rm O/H})+12$.   These lines are produced in the lowest metallicity, highest ionization parameter regions in our starburst models.}
\label{NeIV_pressure}
\end{figure*}

The \NeIV~$\lambda 2425$/\NeIV~$\lambda 2423$ ratio is produced by the ${}^2 D^{0}_{3/2} \rightarrow {}^4 S^{0}_{3/2}$ and the ${}^2 D^{0}_{5/2} \rightarrow {}^4 S^{0}_{3/2}$transitions.   The \NeIV\ lines are only produced at detectable levels (i.e. $1\times 10^{-5}$\Hb) for the lowest metallicity, highest ionization parameter models (i.e. \OH$\leq 7.63$ and $\log(q) \geq 7.75$~dex), as shown in Figure~\ref{NeIV_pressure}.  

\begin{figure*}[!t]
\epsscale{1.0}
\plotone{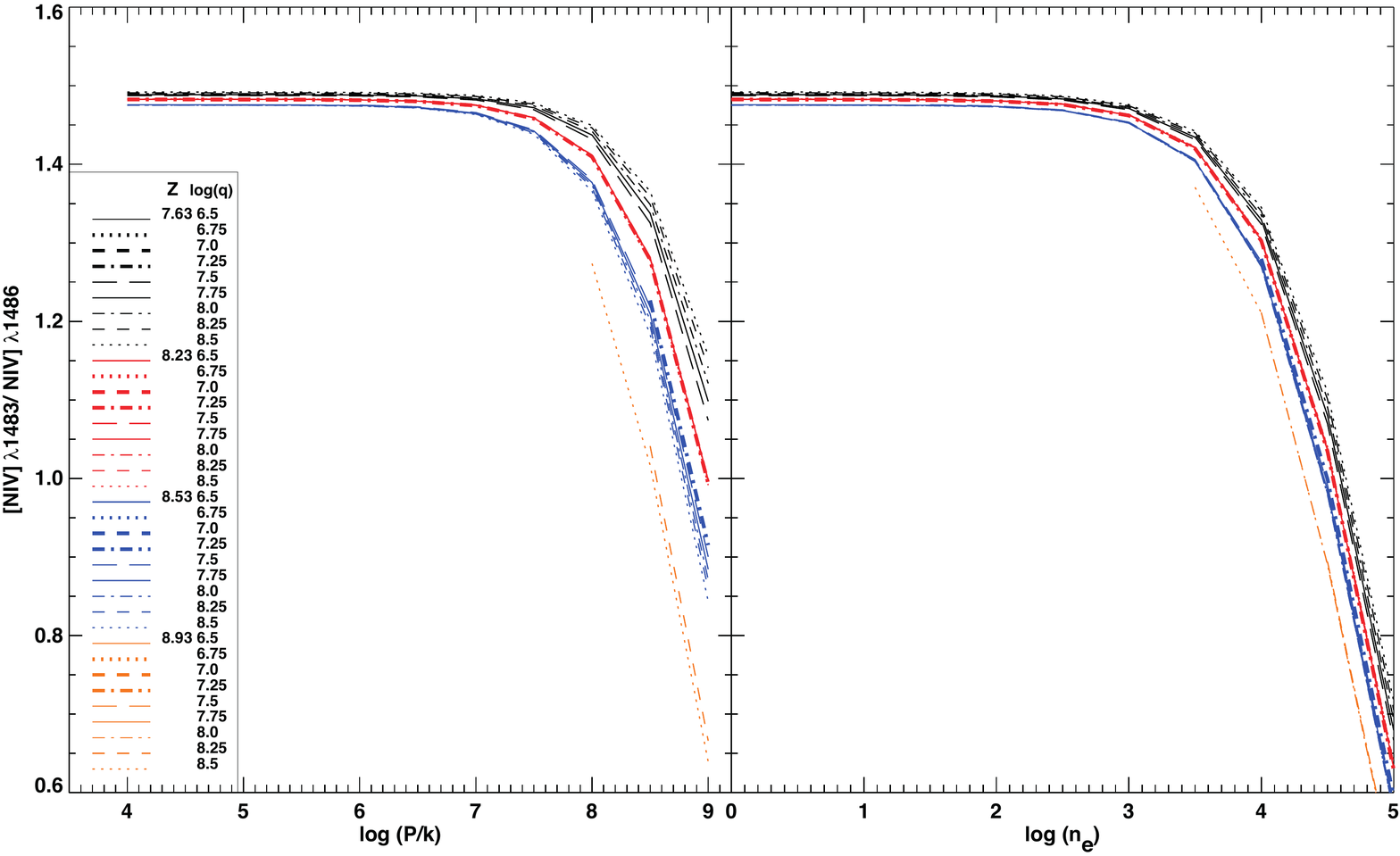}
\caption[fig9.eps]{The theoretical relationship between the \NIV~$\lambda 1483$/\NIVi~$\lambda 1486$ ratio and the ISM pressure (left) and electron density (right) for the metallicities (colored curves) and ionization parameter (solid, dotted, dashed, and dot-dashed lines) covered by our models, as shown in the legend where Z$=\log({\rm O/H})+12$.   These lines are only produced in high temperature, high ionization parameter regions in our starburst models.}
\label{NIV_pressure}
\end{figure*}

The \NIV~$\lambda 1483$/\NIVi~$\lambda 1486$ ratio is produced by the ${}^3 P^0 _2   \rightarrow  {}^1 S_0$ and the ${}^2 P^{o}_{1} \rightarrow {}^1 S_{0}$ transitions.  The \NIV\ and \NIVi\ lines are produced at detectable levels in our high temperature (i.e. low metallicity), high ionization parameter models (i.e. \OH$\leq 8.93$ and $\log(q) \geq 7.5$~dex; Figure~\ref{NIV_pressure}).

Both ratios probe the highest pressure regions of a nebula, and are most commonly observed in regions surrounding AGN, or in gas associated with shocks.  The presence of an AGN or shocks must be ruled out prior to applying our \ion{Ne}{4} and \ion{N}{5} calibrations.

\subsection{Optical Pressure and Density Diagnostics}

There are a large number of ISM pressure diagnostics in the optical.  Many of these diagnostic lines are weak relative to \Hb\ and require deep, high S/N spectra to observe.  Here, we calibrate the traditional electron density diagnostic lines \SII\ and \OII\ with ISM pressure, in addition to the weak optical lines.

\subsubsection{The \OII\ ratio}

The \OII~$\lambda 3729/$\OII~$\lambda 3726$ ratio has been used for several decades to determine the electron density of the gas in planetary nebulae, \HII\ regions, and galaxies \citep[see e.g.,][and references therein]{Pradhan06}.  This doublet is produced by the $ {}^{2} {\rm D}^{0}_{3/2} \rightarrow {}^{4} {\rm S}^{0}_{3/2}$ and $^{2} {\rm D}^{0}_{5/2} \rightarrow {}^{4} {\rm S}^{0}_{3/2}$  transitions.  At temperatures typical of \HII\ regions ($T_{e}=1-2 \times 10^4$~K), the difference in excitation potential between the two upper D levels and the lower S level is $\sim k T$ and therefore the \OII\ transitions are sensitive to the collisional excitation and de-excitation rates, which depend critically on the ISM pressure (or electron density) of the gas.  

 The \OII\ lines have similar ionization potentials to Hydrogen, and are readily observable in optical spectra of \HII\ regions and galaxies, although the close proximity of the two lines requires high resolution spectroscopy to resolve the doublet.
The \OII\ doublet is observable in the optical to $z \sim  1.5$ and is available in the near-infrared (K-band) to $z \sim 5$, and is used to measure the electron density in increasing numbers of galaxies at high redshift \citep[e.g., ][]{Kaasinen16}.  

\begin{figure*}[!t]
\epsscale{1.0}
\plotone{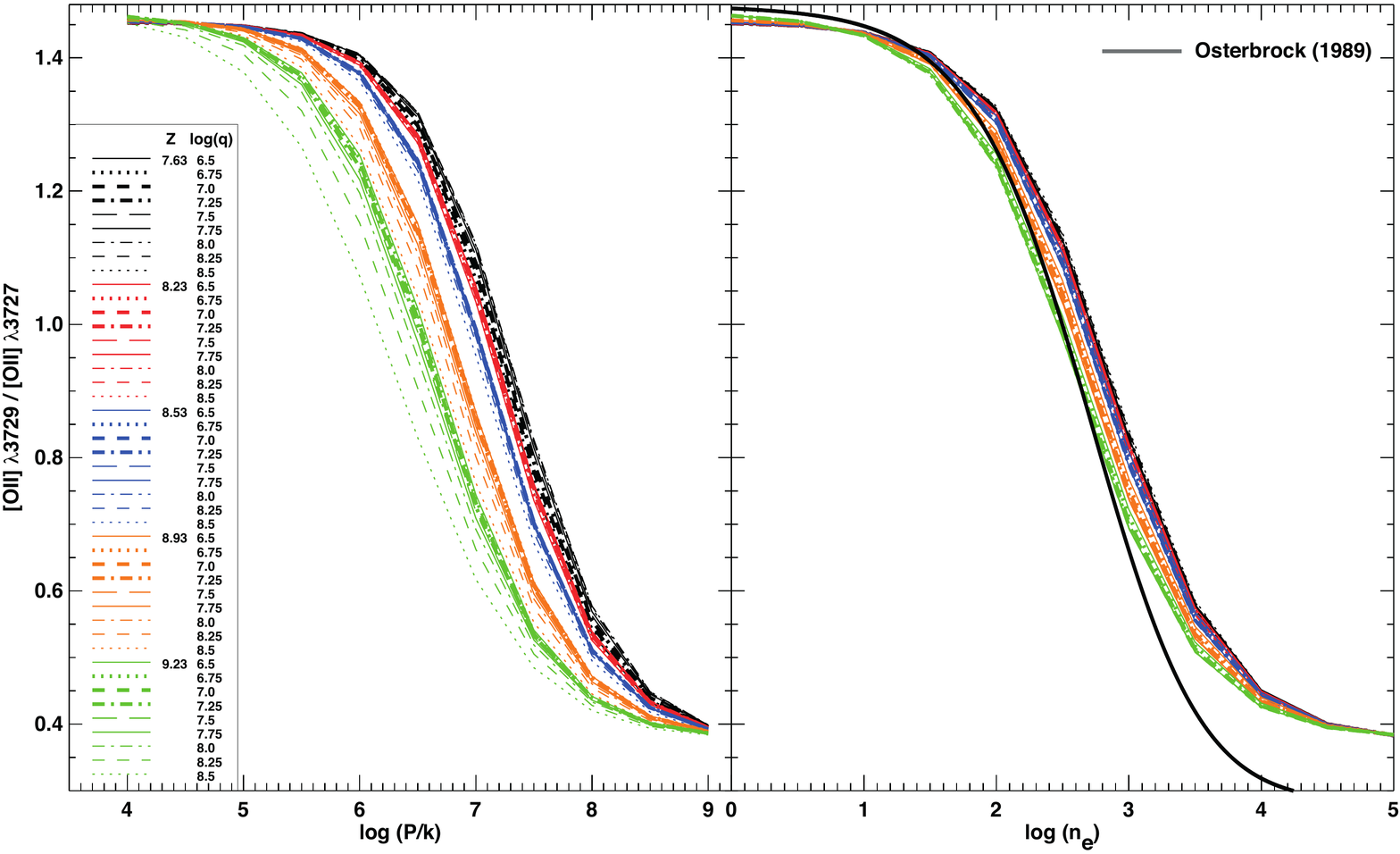}
\caption[fig10.eps]{The theoretical relationship between the \OII~$\lambda 3729/$\OII~$\lambda 3727$ ratio and the ISM pressure (left) and electron density (right) in $cm\,s^{-1}$ for the metallicities (colored curves) and ionization parameter (solid, dotted, dashed, and dot-dashed lines) covered by our models, as shown in the legend where Z$=\log({\rm O/H})+12$.  The relationship between the \OII\ ratio and the ISM pressure depends on the metallicity and the ionization parameter of the gas.  The \OII\ electron density relation from \citep{Osterbrock89} (thick solid black line) is shown for comparison.}
\label{OII_pressure}
\end{figure*}

In Figure~\ref{OII_pressure}, we show how the \OII\ ratio varies with ISM pressure and electron density, for the metallicities and ionization parameters covered by our models.  The relationship between the \OII\ doublet and ISM pressure depends on metallicity (up to 0.4 dex), and there is a secondary dependence on ionization parameter which becomes largest (up to 0.2 dex) at the highest metallicities (\OH$> 8.9$).  

The \OII\ ratio can be used to determine the electron density, with errors of up to 0.4 dex due to the residual dependence of the \OII\ ratio on the electron temperature.  Recall that the electron temperature is sensitive to the metallicity of the nebula.  Therefore, a nebula with the same electron density can give rise to different \OII\ ratios, depending on the metallicity of the gas.  For electron densities more accurate than 0.4 dex, the metallicity should be calculated using consistent strong line calibrations to determine which \OII-density calibration should be applied.

The \OII\ electron density relation from \citet{Osterbrock89} (thick solid black line) is shown for comparison in Figure~\ref{OII_pressure}.  The Osterbrock \OII\ relation has a different shape to our calibration because the atomic datasets for oxygen have changed substantially over the past two decades \citep{Proxauf14}.  

\subsubsection{The \ArIV\ ratio}

The \ArIV~$\lambda 4711,40$ doublet is weak ($\sim 1 \times 10^{-4} \times$\Hb)  for ionization parameters and metallicities typical of star-forming galaxies.  
The \ArIV\ lines are produced by the $ {}^{2} D^{0}_{5/2} \rightarrow  {}^{4} S^{0}_{3/2}$ and  
${}^{2} D^{0}_{3/2} \rightarrow {}^{4} S^{0}_{3/2}$  transitions.  The relationship between \ArIV\ and electron density was first calibrated theoretically by \citet{Stanghellini89}, and updated by \citet{Proxauf14}.  The observed relationship between \ArIV\ and electron density has been investigated primarily for planetary nebulae \citep[e.g.,][and references therein]{Meatheringham91,Copetti02,Wang04}.   

The \ArIV\ lines are produced in thin hot regions of the nebula that are optically thin to UV photons because the ionization potential of the \ArIV\ ratio is large (40.73 eV).  Therefore,  the  \ArIV\ ratio traces a similar ionization zone to the Auroral \OIII$~\lambda 4363$ line, and can be used to probe the ISM pressure where the \OIII$~\lambda 4363$ Auroral line is produced. 
However, care must be taken when measuring the \ArIV\ ratio because the \ArIV~$\lambda 4711$ line is very close to the \HeI~$\lambda 4713$ emission-line and may be blended.  

\begin{figure*}[!t]
\epsscale{1.0}
\plotone{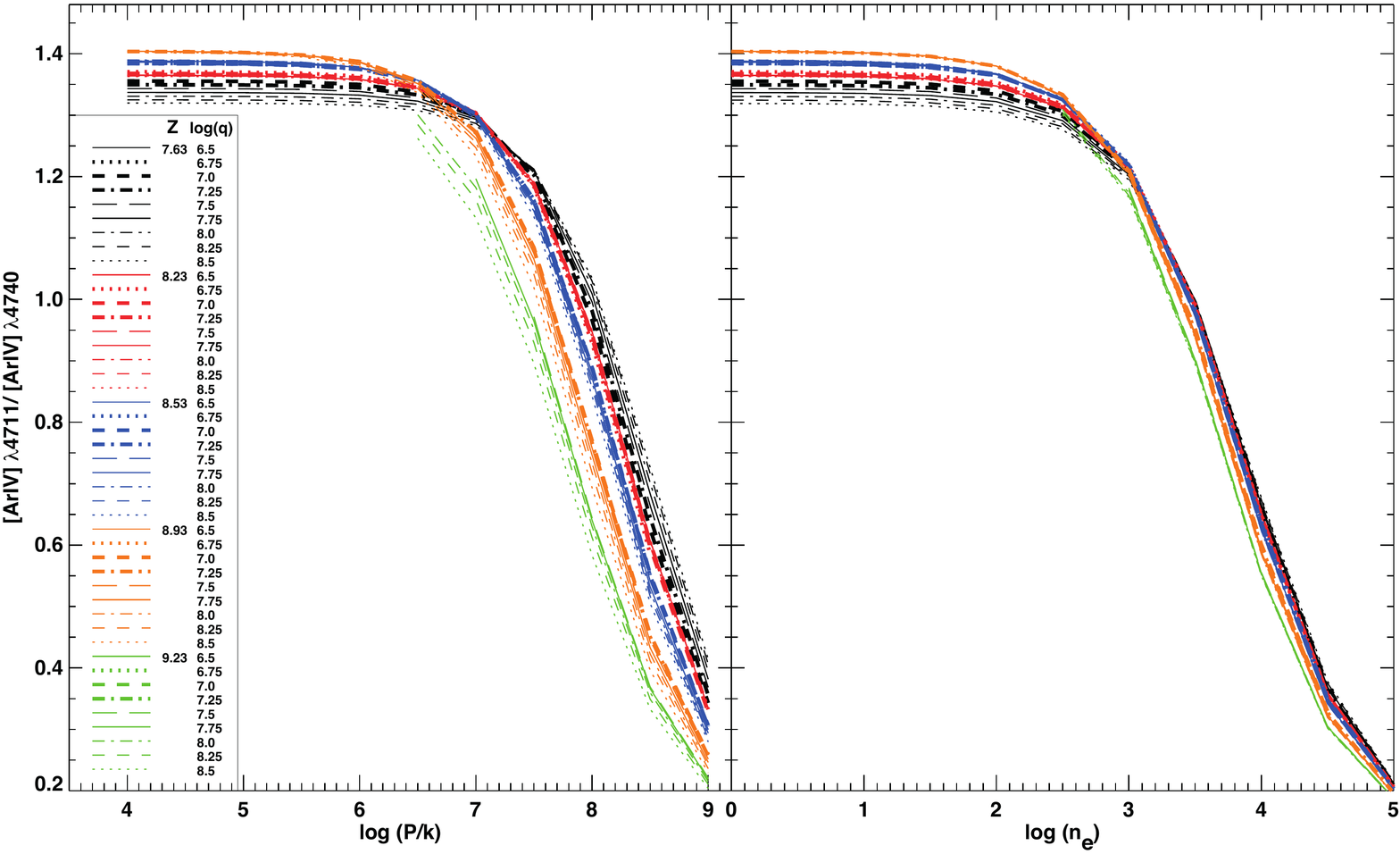}
\caption[fig11.eps]{The theoretical relationship between the \ArIV~$\lambda 4711$/\ArIV~$\lambda 4740$ ratio and the ISM pressure (left) and electron density (right) for the metallicities (colored curves) and ionization parameter (solid, dotted, dashed, and dot-dashed lines) covered by our models, as shown in the legend where Z$=\log({\rm O/H})+12$.  The relationship between the \ArIV\ ratio and the ISM pressure depends on the gas-phase metallicity.  This ratio probes high pressure zones in nebulae.}
\label{ArIV_pressure}
\end{figure*}

Figure~\ref{ArIV_pressure} (left) confirms that the \ArIV\ ratio traces the ISM pressure in the high ionization zone close to the stellar ionizing sources, and is sensitive to pressures above $ \log({P/k}) > 7$.  This ratio is less affected by collisional de-excitation than the other optical pressure-sensitive lines, and is less sensitive to ionization parameter variations (ionization parameter causes the ISM pressure to vary within 0.2 dex for this ratio).  

The  \ArIV\ ratio is a highly sensitive density diagnostic for nebulae with constant electron density; the variation of \ArIV\ with electron temperature (or metallicity) is less than 0.2 dex, and the ratio changes by an order of magnitude between $ 3 <\log(n_e)< 5$.

\subsubsection{The \NI\ ratio}

The \NI~$\lambda 5198$/ \NI~$\lambda 5200$ ratio is a result of the $ {}^{2} D^{0}_{3/2} \rightarrow  {}^{4} S^{0}_{3/2}$  and  $ ^{2} D^{0}_{5/2} \rightarrow  {}^{4} S^{0}_{3/2}$ transitions, respectively.
 This ratio probes the conditions in the outer edges of nebulae where Hydrogen is partially ionized.  The \NI\ ratio was first calibrated as a density diagnostic 
 by \citet{Dopita76}.
 The \NI~$\lambda 5198,5200$ doublet is weak ($\sim 1 \times 10^{-4} \times$\Hb) for ionization parameters and metallicities typical of star-forming galaxies, and has been traditionally measured in planetary nebulae \citep[e.g.,][]{Lee13}.  However, with high resolution, high S/N spectroscopy, it is possible to observe this ratio in \HII\ regions \citep[e.g.,][]{Esteban99}, allowing one to probe the \HII\ region outer recombination zones at the interface between \HI\ and \HII\ regions. 

\begin{figure*}[!h]
\epsscale{1.0}
\plotone{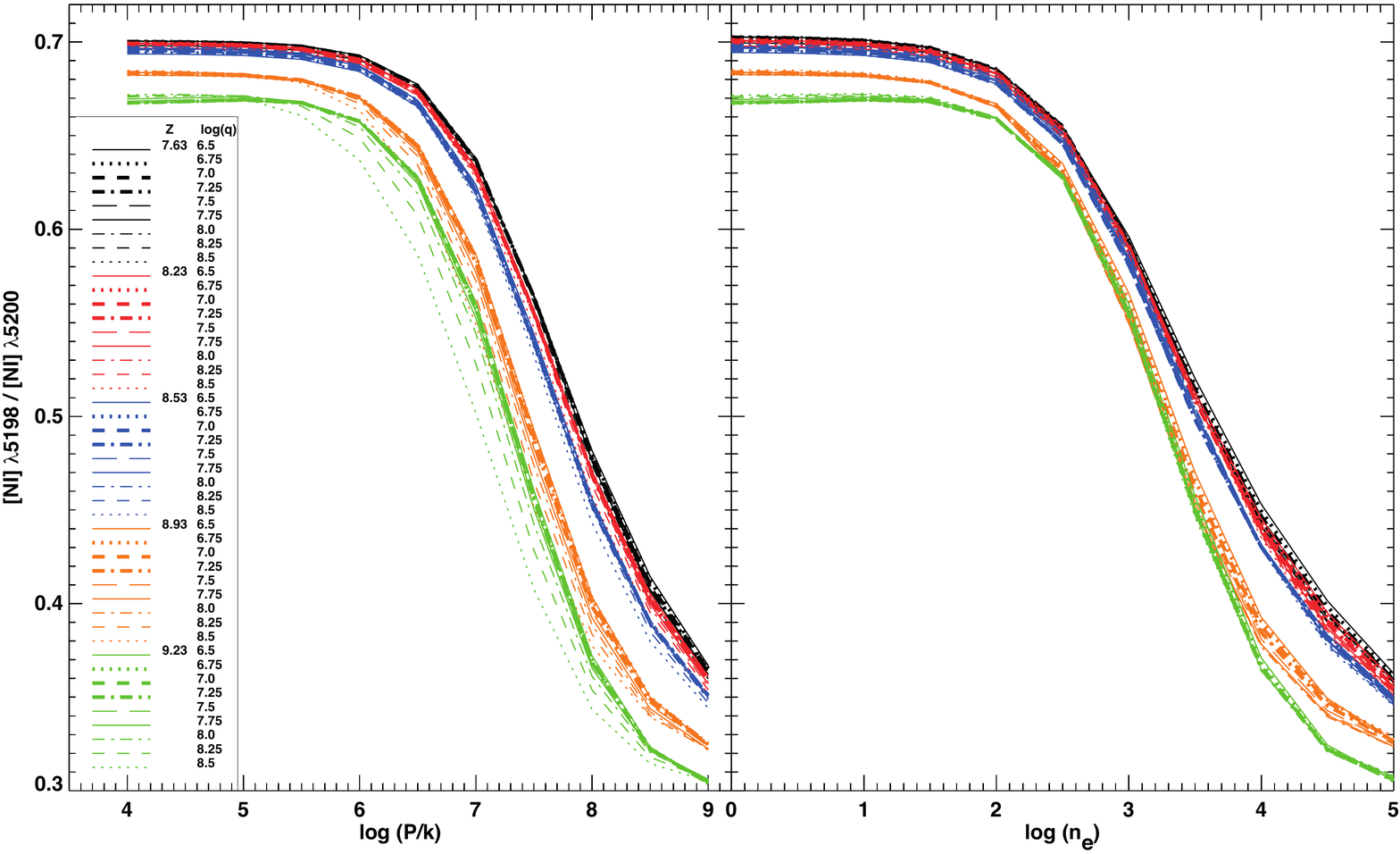}
\caption[fig12.eps]{The theoretical relationship between the \NI~$\lambda 5198,5200$ ratio and the ISM pressure (left) and electron density (right) for the metallicities (colored curves) and ionization parameter (solid, dotted, dashed, and dot-dashed lines) covered by our models, as shown in the legend where Z$=\log({\rm O/H})+12$.  The relationship between the \NI\ ratio and the ISM pressure depends on the metallicity and the ionization parameter of the gas for metallicities above \OH$>8.5$.}
\label{NI_pressure}
\end{figure*}

 Figure~\ref{NI_pressure} (left) shows how the \NI\ ratio depends on ISM pressure as a function of metallicity and ionization parameter.  There is relatively little dependence on metallicity and ionization parameter for \OH$<8.5$ (with differences within $\pm 0.2$~dex).  However, at higher metallicities (\OH$>8.5$), there is a stronger dependence on metallicity and ionization parameter (up to 0.1 and 0.04 dex, respectively).  This behaviour is primarily due to collisional effects between the upper levels in the pair, as well as transitions from higher levels.  
 
Figure~\ref{NI_pressure} (right) shows how the \NI\ ratio depends on the electron density as a function of metallicity and ionization parameter.  While the effect of ionization parameter is minimal ($<0.1$~dex), there is a bimodal dependence on the metallicity, with one curve for low metallicity (\OH$<8.5$), and different curves for high metallicity (\OH$>8.5$).  The saturation at the low density limit occurs at different \NI\ ratios due to multi-level effects and depends on the electron temperature and metallicity.  

\subsubsection{The \ClIII\ ratio}

The \ClIII~$\lambda 5517/$\ClIII~$\lambda 5537$ ratio is produced by the ${}^2 D^{0}_{5/2} \rightarrow {}^4S^{0}_{3/2} $ and ${}^{2} D^{0}_{3/2} \rightarrow {}^4 S^0_{3/2} $ transitions and traces high pressure regimes ($\log({P/k}) > 7$).  The potential use of \ClIII\ as a density diagnostic was first identified by \citet{Weedman68}.  The first calibration of the \ClIII\ ratio in terms of electron density was made by \citet{Aller70}, with subsequent revisions by \citet{Saraph70} and \citet{Keenan92}.  The \ClIII\ ratio has primarily been used to measure the electron density in planetary nebulae \citep[e.g.,][]{Dopita75,Kaler78,Stanghellini89}.  

\begin{figure*}[!t]
\epsscale{1.0}
\plotone{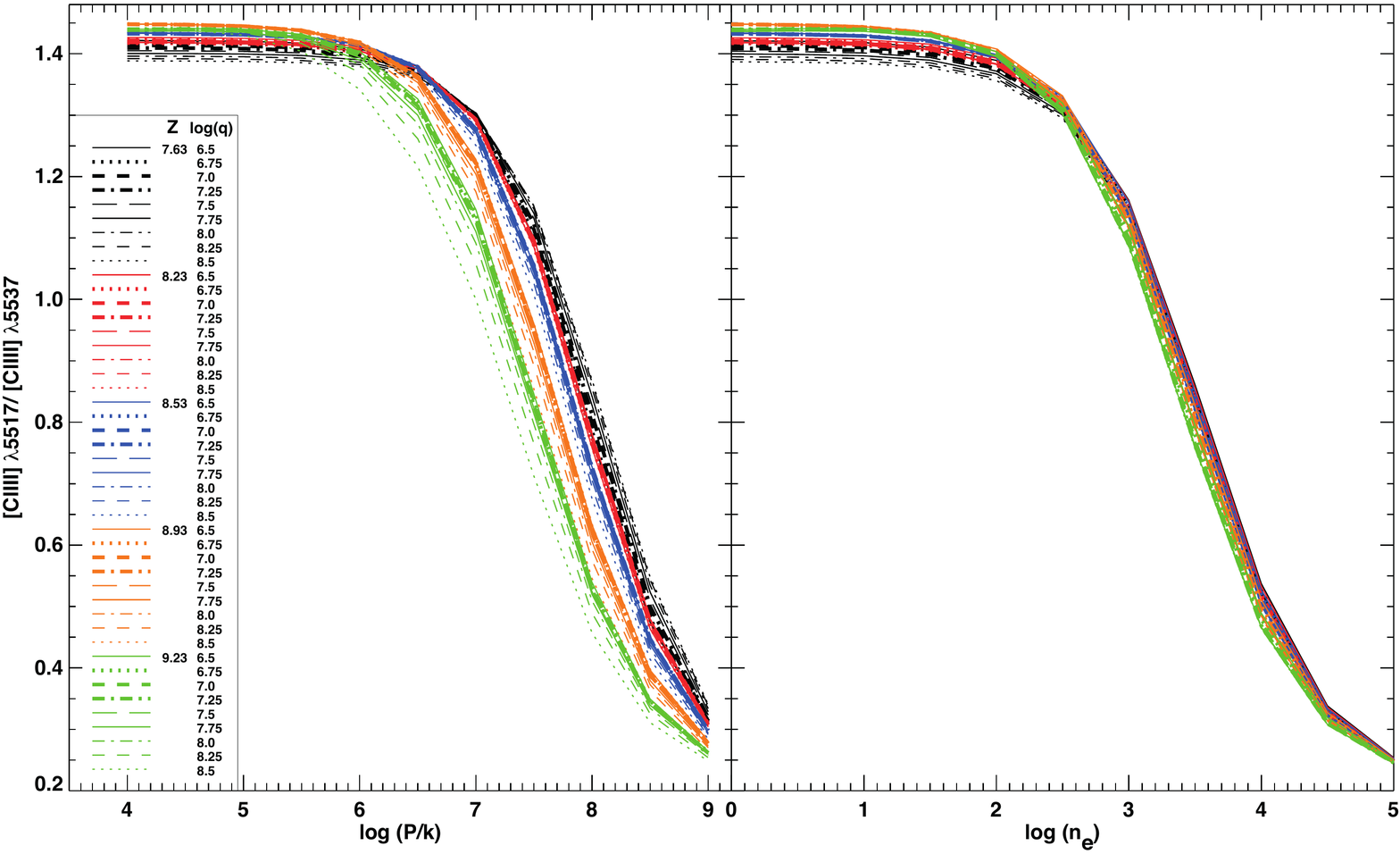}
\caption[fig13.eps]{The theoretical relationship between the \ClIII~$\lambda 5517/$\ClIII~$\lambda 5537$ ratio and the ISM pressure (left) and electron density (right) for the metallicities (colored curves) and ionization parameter (solid, dotted, dashed, and dot-dashed lines) covered by our models, as shown in the legend where Z$=\log({\rm O/H})+12$.  The relationship between the \ClIII\ ratio and the ISM pressure depends primarily on the gas-phase metallicity.  This ratio probes high pressure zones in nebulae.}
\label{ClIII_pressure}
\end{figure*}

Figure~\ref{ClIII_pressure} shows how the \ClIII\ ratio is related to the ISM pressure and electron density.  The relationship between the \ClIII\ ratio and pressure is primarily sensitive to metallicity,  with only minor dependence on the ionization parameter of the gas ($\Delta \log({\rm P/k}) \lesssim 0.3$~dex).  

Like \ArIV, the \ClIII\ ratio is a highly sensitive density diagnostic, with variations $<0.1$~dex due to metallicity for all but the highest metallicities (\OH$>9.2$), where the variation with metallicity is $\sim 0.2$~dex.

\subsubsection{The \SII\ ratio}

The \SII~$\lambda 6731$/\SII~$\lambda 6717$ ratio is produced by the transitions 
$^{2}D^{0}_{5/2} \rightarrow ^{4}S^{0}_{3/2}$ and $^{2}D^{0}_{3/2} \rightarrow ^{4}S^{0}_{3/2}$.   The \SII\ lines have similar ionization potentials to Hydrogen, and are readily observable in optical spectra of \HII\ regions and galaxies.

\begin{figure*}[!t]
\epsscale{1.0}
\plotone{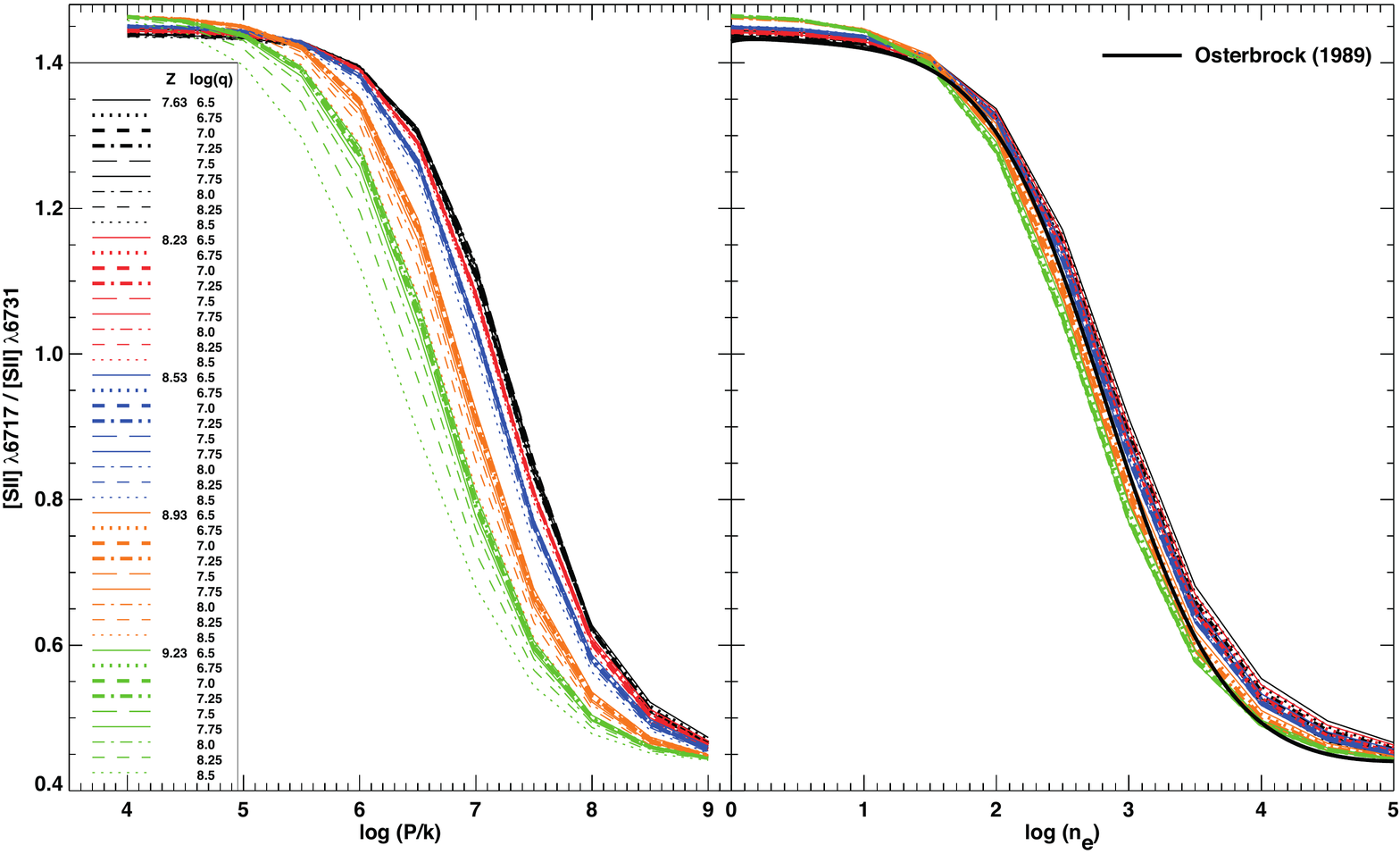}
\caption[fig14.eps]{The theoretical relationship between the \SII~$\lambda 6717,31$ ratio and the ISM pressure (left) and the electron density (right) for the metallicities (colored curves) and ionization parameter (solid, dotted, dashed, and dot-dashed lines) covered by our models, as shown in the legend where Z$=\log({\rm O/H})+12$.  The relationship between the \SII\ line ratio and the ISM pressure depends on the metallicity and the ionization parameter of the gas.  The metallicity can also change the \SII\ density estimate by up to 0.4~dex, for models with constant density. The \SII\ electron density relation from \citep{Osterbrock89} (thick solid black line) is shown for comparison.}
\label{SII_pressure}
\end{figure*}

The theoretical relationship between the \SII\ ratio and ISM pressure depends primarily on the metallicity, with a secondary effect from ionization parameter at metallicities above \OH$>8.9$ (Figure~\ref{SII_pressure}, left panel).  The \SII\ ratio can be used to calculate the ISM pressure up to the \SII\ critical density of $\sim  3 \times 10^3 {\rm cm}^{-3}$, above which the lines are collisionally de-excited.  

The relationship between the \SII\ ratio and electron density (Figure~\ref{SII_pressure}, right panel) has a smaller, but still important dependence on metallicity. 
The metallicity can change the \SII\ density estimate by up to 0.4~dex.   This metallicity dependence should be taken into account when comparing the electron density of \HII\ regions or galaxies where the metallicity is likely to be different, particularly when comparing electron densities derived in galaxies at different redshifts.  

The \SII\ electron density relation from \citet{Osterbrock89} (thick solid black line) is shown for comparison in Figure~\ref{SII_pressure}.  The Osterbrock \SII\ relation is consistent with our calibration, and agrees with our \OH$\sim 8.93$ curves, except in the high density limit, where the Osterbock relation more closely correlates with our \OH$\sim9.23$ curves.
 
\subsection{Infrared Pressure and Density Diagnostics}

The infrared fine structure lines were first used to measure the electron density in planetary nebulae over a decade ago \citep[e.g.,][]{Liu01}.  These lines are now accessible in local \HII\ regions and galaxies, thanks to the {\it Spitzer} and {\it Herschel} satellites 
\citep{Stacey91,Malhotra01,Luhman03,Brauher08,Fischer10}.  Infrared and sub-millimeter instruments now allow the strong infrared fine structure lines to be detected in high-redshift galaxies, probing the ISM conditions over the past 12 billion years \citep[][]{Maiolino05,Maiolino09,Ferkinhoff10,Sturm10,Stacey10,Ivison10,Ferkinhoff11,Valtchanov11,Spinoglio15}.  
Here, we calibrate the \SIII\ 18 and 33$\mu$m, \OIII\ 88 and $52 \mu$m, and \NII\ $ 122 \mu$m and $205 \mu$m lines for pressure and density.  These lines trace a broad range of ISM pressures and densities, providing a complementary suite of diagnostics that can be used to build a comprehensive picture of the pressure or density structure of the ionized gas at a given redshift.

\subsubsection{The \SIII\ ratio}

\begin{figure*}[!t]
\epsscale{1.0}
\plotone{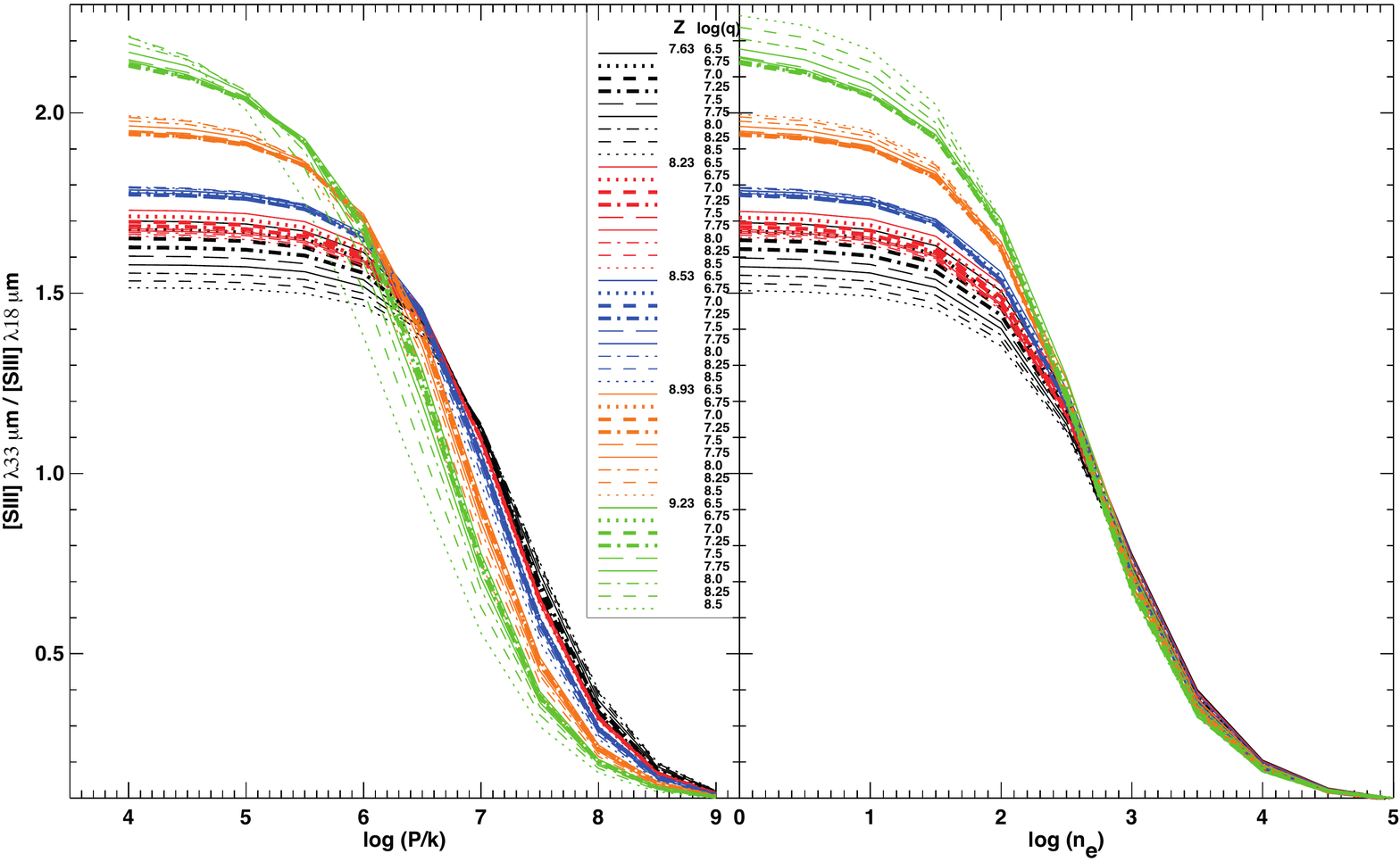}
\caption[fig15.eps]{The theoretical relationship between the \SIII\ $33\mu$m/\SIII~$18 \mu$m ratio and the ISM pressure (left) and the electron density (right) for the metallicities (colored curves) and ionization parameter (solid, dotted, dashed, and dot-dashed lines) covered by our models, as shown in the legend where Z$=\log({\rm O/H})+12$.  The relationship between the \SIII\ ratio and the ISM pressure depends primarily on the metallicity.  Note that the saturation level (y-intercept) at low ISM pressures and densities depends strongly on the metallicity also.}
\label{SIII_IR_pressure}
\end{figure*}

The \SIII\ $33\mu$m/\SIII~$18 \mu$m ratio is readily observable in the rest-frame infrared spectrum.  The \SIII\ ratio is produced by the $ ^{3} P_{1} \rightarrow  ^{3} P_{0}$ and $ ^{3} P_{2} \rightarrow  ^{3} P_{1}$ transitions and traces the high ionization, high pressure regions of a nebula ($7<\log({\rm P/k})<8$) as shown in Figure~\ref{SIII_IR_pressure}.  

For all pressures between $6<\log({\rm P/k})<8$, the metallicity dependence is strong (up to an order of magnitude in $\log({\rm P/k})$ across the full metallicity range of our models).  The dependence on ionization parameter is only important in the most metal-rich nebulae (\OH$>9.2$).  At low pressures, the saturation value depends strongly on the metallicity, with a strong dependence on the ionization parameter at the lowest metallicities (\OH$<8.0$) due to multi-level effects.

Figure~\ref{SIII_IR_pressure} indicates that the infrared \SIII\ ratio is a sensitive diagnostic of the electron density for high density environments ($2.5<\log ( \frac{n_e}{cm^3} ) <4$).  In this regime, there is no residual dependence on metallicity or ionization parameter.  However, below $\log ( \frac{n_e}{cm^3} ) <2.5$, the \SIII\ ratio is much more sensitive to metallicity than the electron density, and the saturation value is also metallicity dependent.

\begin{figure*}[!t]
\epsscale{1.0}
\plotone{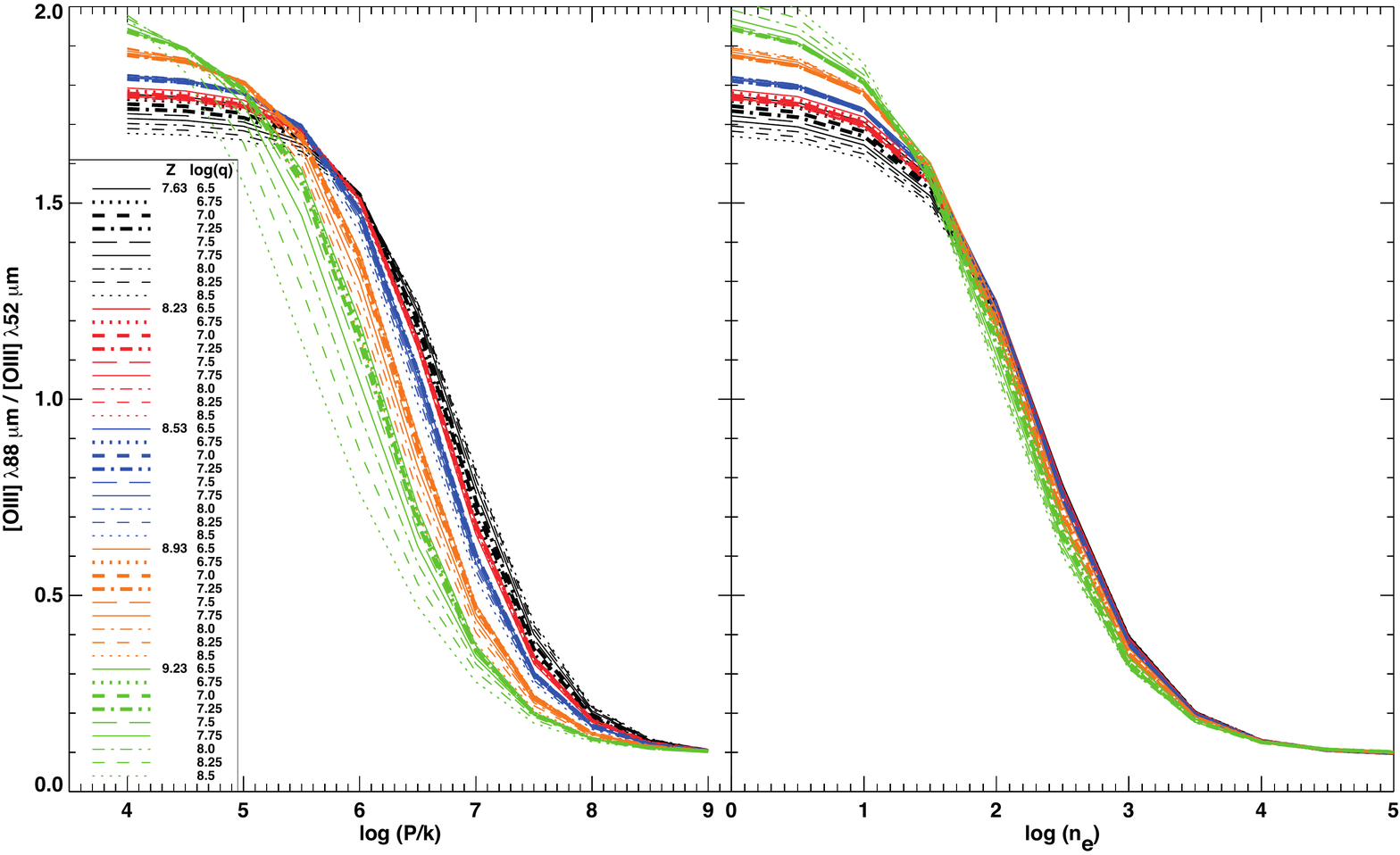}
\caption[fig16.eps]{The theoretical relationship between the \OIII~$88 \mu$m/\OIII~$52 \mu$m ratio and the ISM pressure for the metallicities (colored curves) and ionization parameter (solid, dotted, dashed, and dot-dashed lines) covered by our models, as shown in the legend where Z$=\log({\rm O/H})+12$.  The relationship between the \OIII\ ratio and the ISM pressure depends on the metallicity and the ionization parameter of the gas.}
\label{OIII_IR_pressure}
\end{figure*}

\subsubsection{The \OIII\ ratio}

The {\it Herschel} Observatory has allowed the \OIII~$88 \mu$m/\OIII~$52 \mu$m ratio to be measured for increasing numbers of star-forming galaxies in the local universe, and is now also possible to observe the \OIII\ lines in galaxies at high redshift \citep{Ferkinhoff10}.  These lines probe \HII\ regions with densities up to $300\, {\rm cm}^{-3}$, and are produced by the transitions  $ ^{3} P_{2} \rightarrow  ^{3} P_{1}$ and $ ^{3} P_{1} \rightarrow  ^{3} P_{0}$. 

The \OIII\ ratio is strongly affected by metallicity, with the \OIII-pressure calibration varying by up to an order of magnitude in $\log({\rm P/k})$ from \OH$=7.6$ to \OH$=9.23$ (Figure~\ref{OIII_IR_pressure}, left panel).  The dependence on ionization parameter is $\sim 0.2$~dex for metallicities \OH$<8.9$, rising to 0.5 dex above \OH$>8.9$.   At low pressures, the \OIII\ ratio saturation value depends on the metallicity.  As for \SIII, the stronger metallicity (i.e. temperature) sensitivity at low density is due to multi-level effects.  

The relationship between the \OIII\ ratio and electron density is largely independent of metallicity (to within 0.1 dex) except in the most metal-rich galaxies (\OH$>9.2$) where there is a 0.2 dex offset from the \OIII-electron density relationship.  Therefore, the \OIII\ ratio is a robust indicator of the electron density for electron densities $>100$~cm$^{-3}$, below which the \OIII\ ratio enters the low density limit.

\subsubsection{The \NII\ ratio}

The infrared fine structure lines of \NII~$205 \mu$m and \NII~$122 \mu$m are produced by the $^3 P_{1}  \rightarrow ^{3} P_{0}$ and the $^{3} P_{2} \rightarrow ^{3} P_{1}$ transitions.  These lines have a low ionization potential (14.5~eV) and are produced within \HII\ regions around late O-type and early B-type stars.
They are sensitive to the low to intermediate pressure regime ($4<\log({\rm P}/k)<6.5$), providing a complementary diagnostic to the IR \SIII\ and \OIII\ ratios.  With {\it Herschel}, the \NII\ ratio provides estimates of the electron density in both galaxies \citep{Zhao16,Herrera-Camus16} and \HII\ regions \citep{Goldsmith15}.  Sub-mm observations now allow the \NII\ lines to be detected in high redshift galaxies \citep{Ferkinhoff11,Decarli12,Ferkinhoff15}.

\begin{figure*}[!t]
\epsscale{1.1}
\plotone{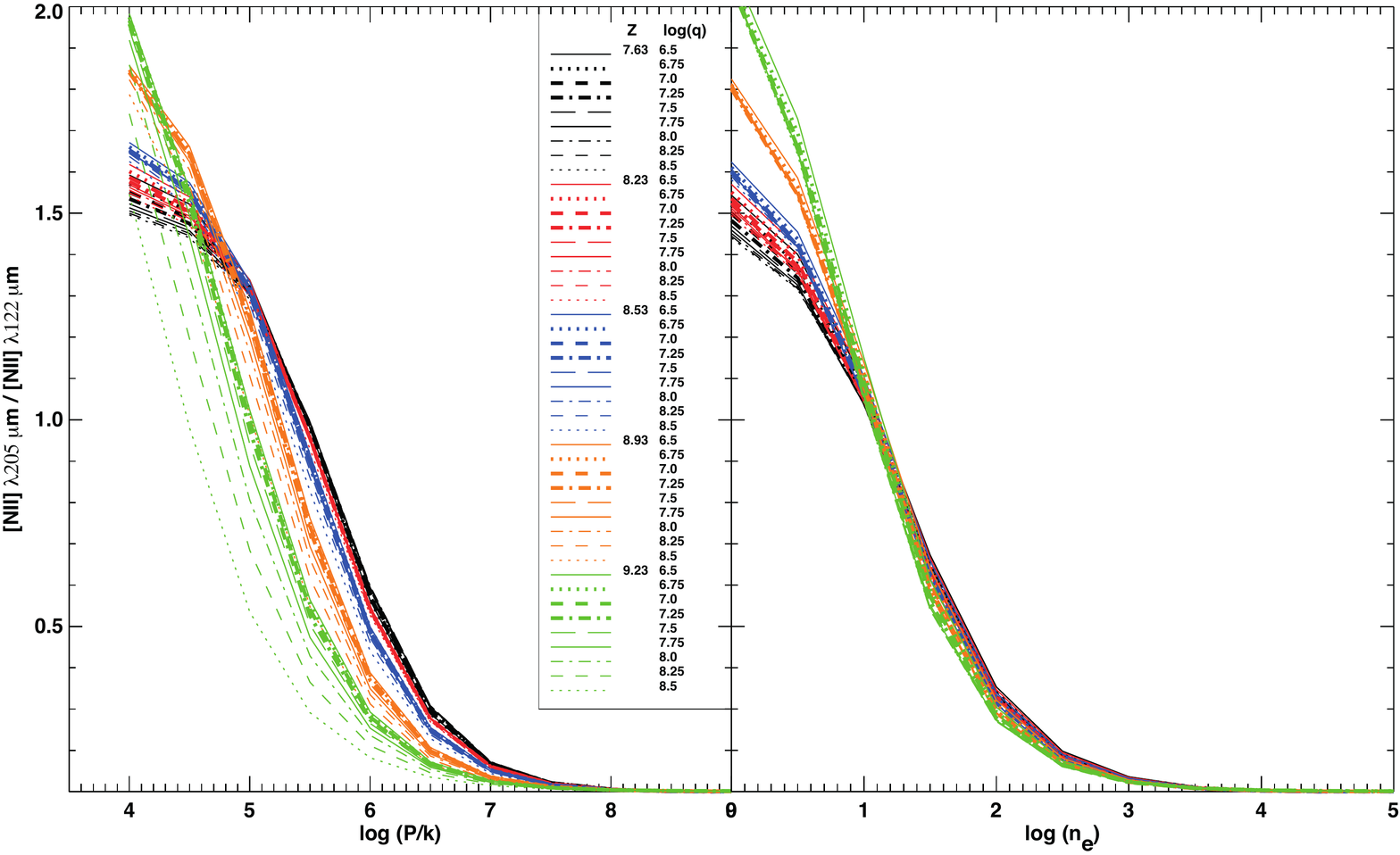}
\caption[fig17.eps]{The theoretical relationship between the \NII~$\lambda 205 \mu$m/\NII~$\lambda 122 \mu$m ratio and the ISM pressure (left) and electron density (right) for the metallicities (colored curves) and ionization parameter (solid, dotted, dashed, and dot-dashed lines) covered by our models, as shown in the legend where Z$=\log({\rm O/H})+12$.  The relationship between the \NII\ ratio and the ISM pressure depends on both the metallicity and the ionization parameter of the gas.}
\label{NII_IR_pressure}
\end{figure*}

Figure~\ref{NII_IR_pressure} shows that the \NII\  ratio is extremely sensitive to the ISM pressure, varying by $\sim 1.5$ orders of magnitude across the pressures probed by our models.   The \NII\ ratio is also sensitive to the gas-phase metallicity.  The ISM pressure can vary with metallicity by up to $\sim 1.5$~dex for $\log({\rm P}/k)<5$), depending on the ionization parameter.  The \NII\ ratio remains sensitive to both the ISM pressure and metallicity even at the lowest pressures ($\log({\rm P}/k)<4$).  Therefore, the \NII\ ratio can provide a powerful probe of the ISM pressure across a broad range of potential pressures, provided that the gas-phase metallicity is known.  For the most metal-rich \HII\ regions and galaxies (\OH$>9.0$), the ISM pressure calibration also depends on the ionization parameter of the gas ($\pm 0.2$~dex) and calculation of the ISM pressure should include the effect of ionization parameter as well as metallicity.

The \NII\ ratio is a robust diagnostic of the electron density for a large range of possible electron densities ($1<\log({\rm P}/k)<2.5$).  The \NII\ ratio can also potentially be used to calculate the electron density in low density regimes ($\log({\rm P}/k)<1$), providing the electron density can be corrected for metallicity.

\section{Discussion}\label{discussion}

Several important issues affect the ISM pressure or electron density derived from the spectra of \HII\ regions or galaxies.
Different species and ionization states of elements will produce different ISM pressure or density estimates, and should not be directly compared unless it is clear that the different species are probing the same zones within the ionized nebula.  The ionization and density structure of the nebula may lead to different species probing different regions of gas within an \HII\ region.  The atomic data used to calculate the line ratio calibrations can also cause intrinsic offsets among ISM pressures or densities of different species, even if those species probe the same regions within a nebula.  In addition, some density-sensitive lines may be contaminated by emission from the diffuse ionized gas, or from shock excitation.  We discuss each of these issues below.

\subsection{Density Structure}

\HII\ regions have extremely complex density structures.  \HII\ regions can trigger the formation of new stars through ``collect and collapse" \citep{Elmegreen95} and ``radiation-driven implosion" \citep{Bertoldi89}.  In the ``collect and collapse" model proposed by Elmegreen et al., the expansion of an \HII\ region into a supersonically turbulent cloud causes coagulation of clumps which are gravitationally unstable.  These unstable clumps can then collapse and form new stars.   In the radiation-driven implosion scenario, hot stars penetrate the ISM and heat the cold low-density gas.  This heating amplifies overdensities created by the turbulent ISM, which can then collapse and form stars \citep{Gritschneder09,Dale12}.  These processes can create globules of dense gas \citep[e.g.,][]{Tremblin13,Walch15,Schneider16}, while stellar winds can drive outflows that create horse-shoe or other complex geometries \citep{Park10}.

Ionized gas or radio continuum density measurements within \HII\ regions show many complex radial gradients \citep{Franco00,Perez01,Binette02,Phillips07,Herrera-Camus16} and in some cases little to no gradient \citep{Ramos-Larios10,Garcia-Benito10}.  The density gradients in \HII\ regions are related to \HII\ region size.  Ultracompact \HII\ regions often show steep density gradients \citep{dePree95,Franco00,Kurtz02,Johnson03b,Phillips07}, while larger \HII\ regions typically have more shallow or flat density gradients \citep{Phillips08}.   Density gradients can be approximated by $n \propto r^{-\beta}$ where $n$ is the density and $r$ is the distance of the gas from the central stellar source.  \citet{Franco00} found steep gradients in three ultra-compact \HII\ regions with exponents $1.6<\beta<2.4$.  On the other hand, in normal Galactic \HII\ regions, large-scale density gradients appear to be relatively modest. \citet{Phillips08} found that $\beta \leq 1$ and that density variations are small ($\Delta n \sim 10^2 {\rm cm}^{-3}$ on average). 

The density structure of nearby Seyfert galaxies has recently been measured by \citet{Spinoglio15} using {\it Herschel} data.  They find an anti-correlation of the electron temperature with the gas density, and they find significant density stratification within the galaxies.  On average, electron densities increase with the ionization potential of the ions.  In a clumpy nebula with high density gas clumps, the clumps are likely to be unresolved with standard integral field or integrated spectroscopy.  A clumpy ISM has also been observed in high redshift galaxies \citep{Carniani17}.

Where density gradients or complex density structures are expected, we recommend the use of constant pressure diagnostics rather than constant density diagnostics.  Our pressure models allow a unique density to be calculated at each step through a nebula, and can produce density gradients or other profiles.  If our constant density diagnostics are applied to \HII\ regions or galaxies that contain density gradients or clumps, the density derived will likely be dominated by the regions that produce the largest emission-line strengths, and will not necessarily represent an average electron density for the nebulae.

Emission-lines produced by different ions and different energy levels are sensitive to different density regimes, depending on the critical density of the transition.  Lines with low critical densities, like \OII\ and \SII, are affected by collisional de-excitation and are therefore weak in the high density regions, while \CIII, \ArIV\  trace higher density regions of a nebula.  The \OIII\ fine structure lines have a significantly lower critical density than the \ArIV\ or the \ClIII\ lines, and therefore trace lower density regions.  In addition, where the density varies through a nebula, the density estimated from diagnostics from species predominantly in the outer regions of the nebula, such as the \SII\ or \NI\ ratios, may not give reliable estimates of the density in regions where the electron temperature has been calculated (typically the \OIII\ lines) (Nicholls et al., in prep). Where possible, co-spatial species should be used.  In the case of \OIII, the the \ClIII\ density diagnostic may be a better choice, if observed.

\begin{figure*}[!t]
\epsscale{1.0}
\plotone{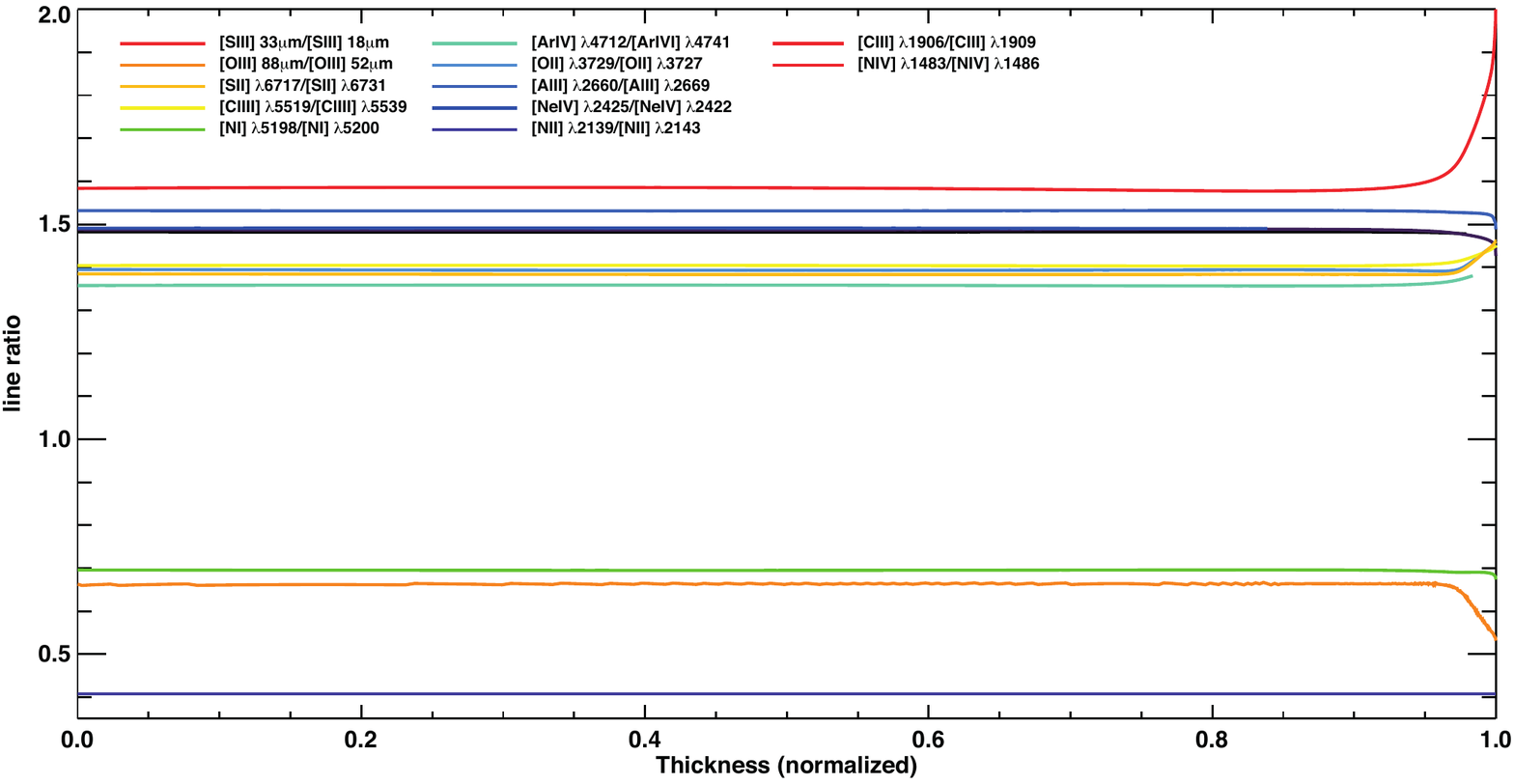}
\caption[fig18.eps]{Normalized thickness (radius) versus emission-line ratios for pressure and density sensitive emission-line ratios in a spherical model with \OH$=8.23$, $\log(q)=8.0$~cm/s, and $\log({\rm P/k})=6$~dyn.  Coloured curves correspond to different emission-line ratios, as shown in the legend}
\label{ratios_radius}
\end{figure*}

In Figure~\ref{ratios_radius}, we show how the line ratios vary as a function of radius throughout a nominal spherical model with \OH$=8.23$, $\log(q/cms^{-1})=8.0$~cm/s, and $\log({\rm P/k})=6$.  The line ratios are relatively constant with radius (i.e. constant to within $\pm 0.01$~dex, until the partially ionized zone that occurs in the last $\sim10$\% of the nebula).

\begin{figure}[!t]
\epsscale{1.2}
\plotone{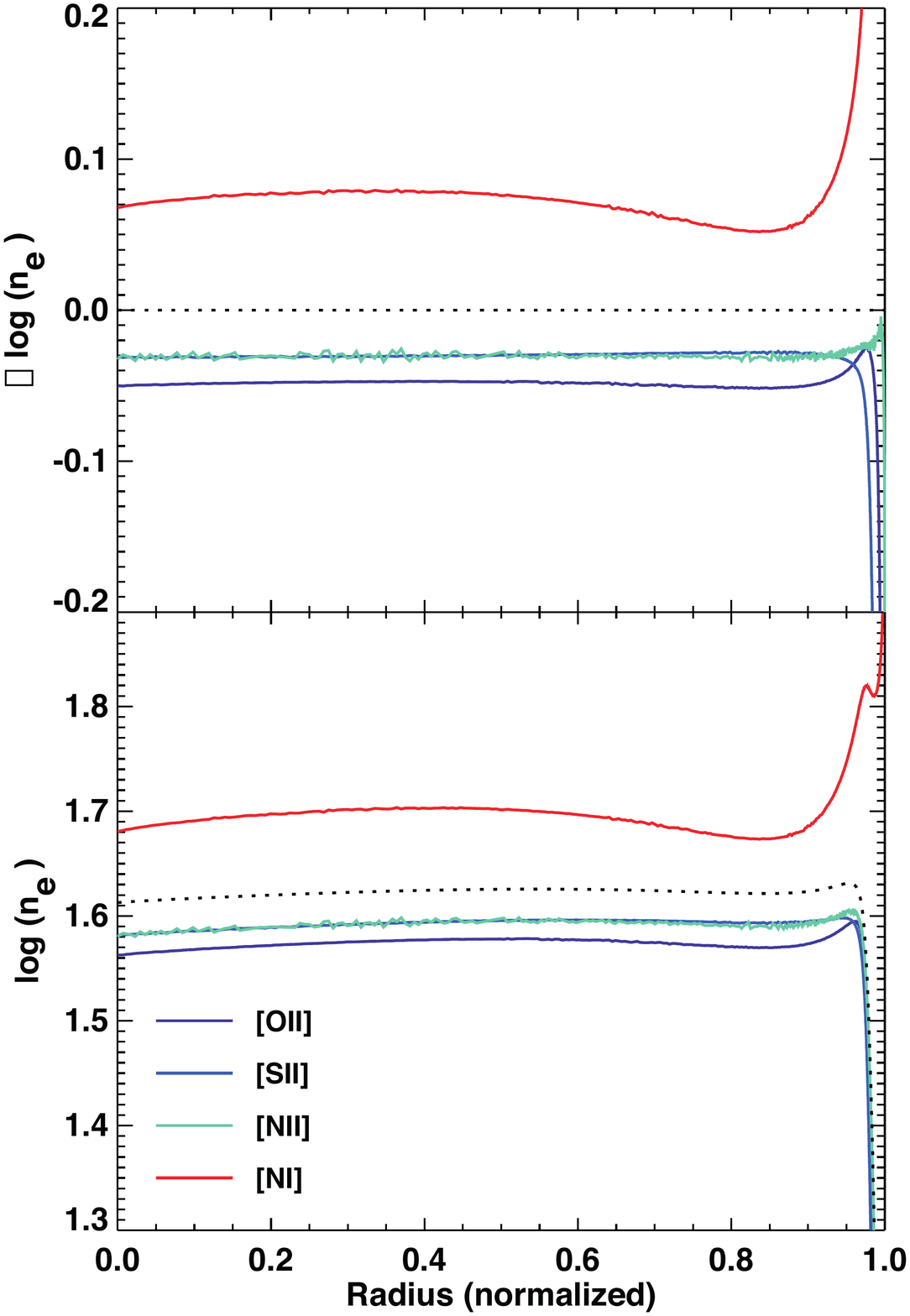}
\caption[fig19.eps]{Normalized thickness (radius) versus the logarithm of the electron density for the  \SII~$\lambda 6717/$\SII~$\lambda 6731$, \OII~$\lambda 3729/$\OII~$\lambda 3727$, \NII~$205\mu$m/\NII~$122\mu$m, and \NI~$\lambda 5198/$\NI~$\lambda 5200$ emission-line ratios in a spherical model with \OH$=8.23$, $\log(q)=8.0$, and $\log({\rm P/k})=6$.  The black dashed line shows the true density of the models.  The \OII\ and \NII\ ratios provide the best fit to the true density; these ratios agree with the true density to within 3\%.  The upper panel shows the relative offset among the different density indicators.}
\label{radial_density}
\end{figure}

To test how well our density diagnostics (which have been derived using plane parallel models) could be used to map the density structure of a roughly spherical HII region using integral field observations, we use our spherical pressure models and derive the density at each radial step throughout the nebula for our nominal spherical model.   This test is based on the assumption that the \HII\ region is roughly spherical, is well resolved, that the metallicity and ionization parameter of the \HII\ region are known, and that the \HII\ region does not contain any additional power sources such as shocked ionizing radiation fields from supernovae or stellar winds.  However, this model is useful because for a pressure of $\log({\rm P/k})=6$, the electron density is constant at $\sim 40$~cm$^{-3}$ throughout the nebula until the gas begins to recombine at the edge of the nebula.  An electron density of $\sim 40$~cm$^{-3}$ is close to (but not within) the low density limit of the commonly-used \SII\ and \OII\ doublets, and provides a useful test of how well our models are able to reproduce the density within this regime, which is difficult to fit analytically.  Figure~\ref{radial_density} shows how well the electron diagnostics using \SII~$\lambda 6717/$\SII~$\lambda 6731$, \OII~$\lambda 3729/$\OII~$\lambda 3727$, \NI~$\lambda 5198/$\NI~$\lambda 5200$, and \NII~$205\mu$m/\NII~$122\mu$m track the electron density through the spherical nebula.  The \SII\ and \NII\ ratios provide the best match to the true density throughout the nebula, within 3\% on average.  The \OII\ and \NII\ ratios agree with the true density to within 5\% and 7\% on average.   
We emphasize that these electron densities have been derived using plane parallel diagnostics.  Spherical models have a larger partially ionized zone than plane parallel models.  Lines that are produced in the partially ionized zone (such as [NI]) will give the poorest fit to the average density of the spherical model.  This effect will become larger for models with high ionization parameters.  

Modelling a clumpy ISM in nebulae may overcome much of these problems.  Monte Carlo simulations are needed to track individual lines of sight throughout a nebula.  Monte Carlo photoionization models have been developed over the past two decades \citep{Ercolano03, Ercolano05,Ercolano08,Robitaille11,Lomax16}.  These models successfully reproduce properties in individual star forming regions, but the physics included in these models has been constrained by computation time. For example, the models currently don't include clumpy dust distributions, which are modeled with separate dust Monte Carlo codes which have simple photoionization calculations \citep[e.g.,][]{Gordon01,Steinacker13,Gordon17}.  We expect significant advances to be made in this area in the coming years.

\subsection{Ionization structure}

\begin{figure}[!h]
\epsscale{1.2}
\plotone{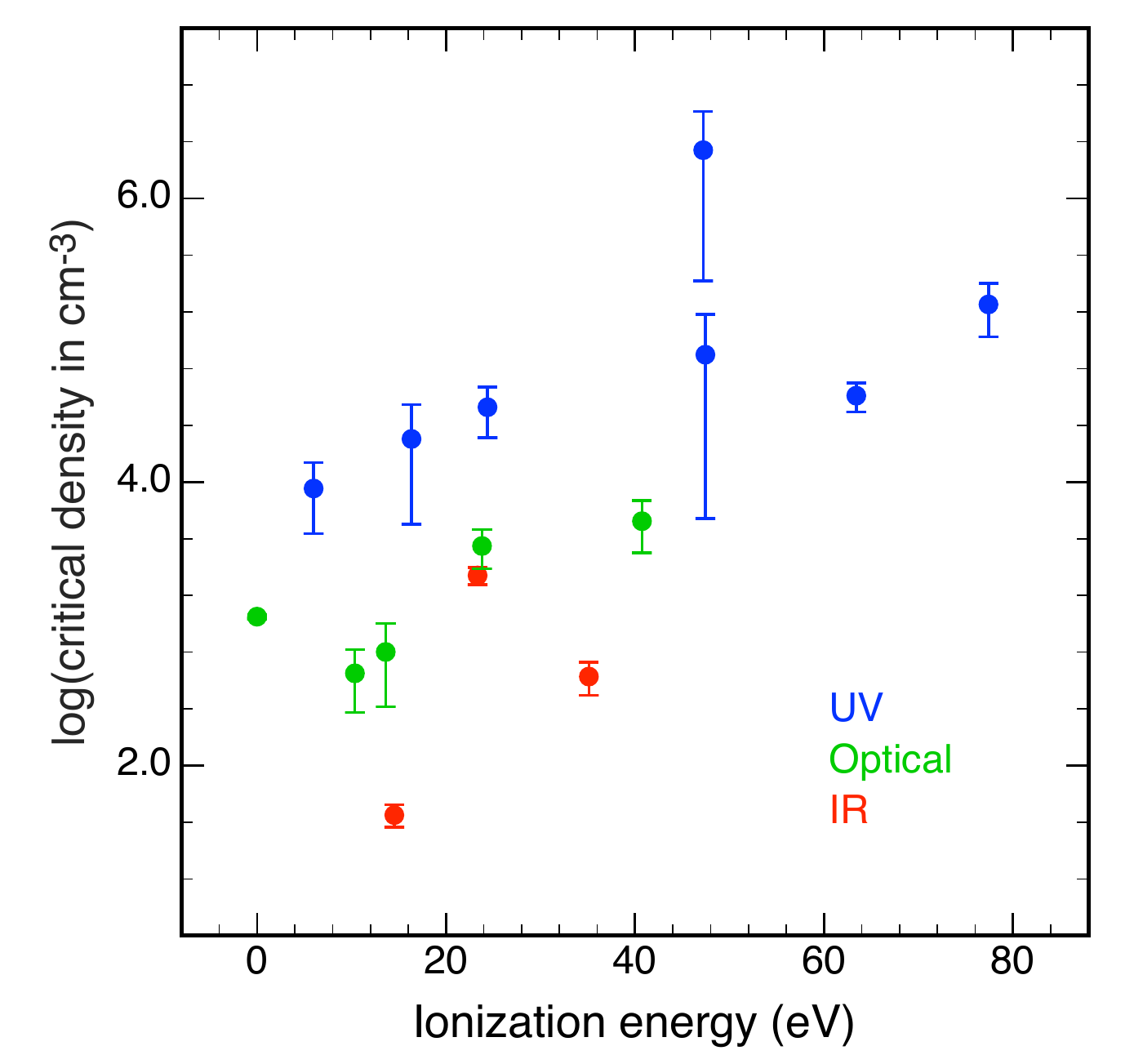}
\caption[fig20.pdf]{The ionization energy and critical densities of the pressure-sensitive species considered in this work.  The ionization energy here is the energy required to create the ion in its ground state, not the energy required to ionize the atom into a higher ionization state. The uncertainties arise from the horizontal spread in $n_e$ in Figures 1,2, and 3 due to temperature variations. The larger the error bars, the greater the temperature dependence of an individual critical density.}
\label{ionization_potential}
\end{figure}

Different ionization lines of different species probe different zones of a nebula.  Thus, by observing several different ionization lines of different species, we can derive a picture of the pressures within the different ionization zones of an \HII\ region.  Figure~\ref{ionization_potential} shows the ionization energy and critical densities of the species considered here, where the ionization energy is the energy required to create the ion in its ground state, not the energy required to ionize the atom into a higher ionization state.  

We use ionization data from NIST \citep{Kramida15}, as given in Table~\ref{table_ne}.  The uncertainties in the critical densities represent the potential variation due to electron temperature (as in our Temperature Models).  The density-sensitive species in the optical and UV cover a wide range of critical densities and ionization energies.

\begin{figure*}[!t]
\includegraphics[width=\linewidth]{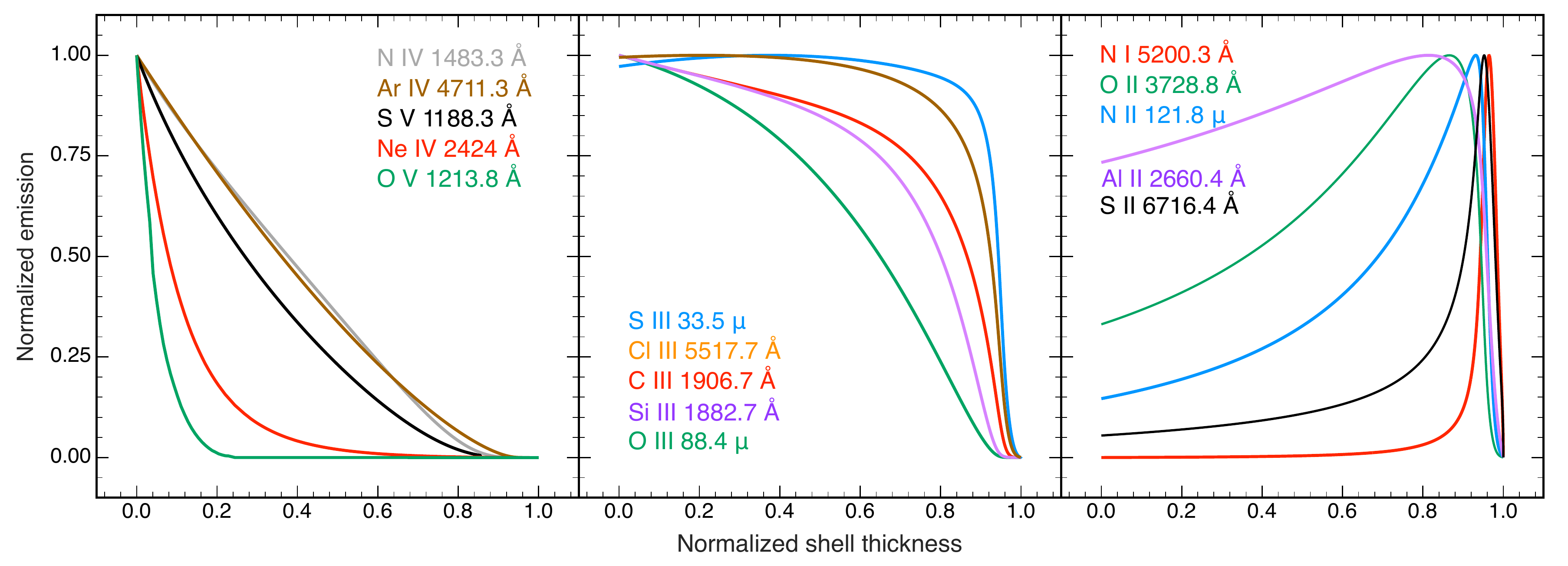}
\caption[fig21.pdf]{The ionization zones for the density-sensitive emission-line pairs for which we provide ISM pressure and density calibrations (described in Section~\ref{pressure}).  The different panels show how different levels of ionization probe different regions of the nebula.}
\label{ionization_zones}
\end{figure*}

In Figure~\ref{ionization_zones}, we show the MAPPINGS v5.1 ionization zones for the species considered here.  The \NeIV, \OV, \ArIV, \NIV, and \SV\ ratios ratios trace regions of high ionization.  Therefore, electron densities derived for these regions will be larger than for the lower ionization lines, if the \HII\ regions have negative density gradients.

The ionization potential of higher levels of ionization (i.e. higher than the ionization state of the ion used for our diagnostics) also play a role in defining the regions where each line is observable.  For example, the ionization potential of the \SI~$\rightarrow$~\ion{S}{2} transition (12.20 eV) is lower than the ionization potential of hydrogen.  Therefore, the \SII\ states can be readily achieved by thermal absorption of Lyman photons from hydrogen recombination or by collisional excitation.  The higher ionization state \SIII\ requires a significantly larger energy to achieve (23.3 eV). This higher \SIII\ level can be achieved in regions that are optically thin to UV photons or from photons from He recombination.  Therefore, the majority of the \SII\ photons are likely to be produced in high pressure clumps or on the outskirts of nebulae, and not in the same zone as the \OII\ lines \citep[see e.g.,][for a discussion]{Proxauf14}.   Figure~\ref{ionization_zones} highlights the fact that the \SII\ lines are produced in the extended partially ionized zone of the nebula, while the \OII\ lines are produced closer to the ionizing source.  Therefore, the \SII\ and \OII\ lines should not be used interchangeably to measure the ISM pressure or density unless the density is constant across the \HII\ regions or galaxies within the sample(s) being studied.

Low ionization species, such as \NI\ can probe the pressure at the very edge of the nebula, where the \HII\ region transitions into an \HI\ region.  Note that some additional pressure-sensitive lines such as the infrared \OI\ and \CI\ lines are produced partially or predominantly in the photo-dissociation region or in neutral gas, which are not tracked by our MAPPINGS v5.1 models.  We refer the reader to \citet{Tielens85} for a description of the use of photodissociation models to derive electron densities in the outer nebula zones.

Only a few sets of lines trace the same region of the nebula. The \ClIII\ and \SIII\ lines are both produced throughout the nebula, and derived pressures may be considered as broadly representative of the overall ISM pressure within the nebula.   Both the \SII\ and \NI\ lines trace the very outer edge of the nebula, while the high ionization species (such as \ArIV, \NV, \NeIV) trace the inner zones close to the ionizing source.   

The \NII\ and \OII\ lines trace similar regions of the nebula, and may be used to trace intermediate-zone pressures at optical and UV wavelengths, respectively.
Therefore, the \NII\ and \OII\ lines can be used to trace or compare the ISM pressure in local and high redshift galaxies.

\subsection{Atomic Data}

The pressures or densities derived for different species are critically dependent on the atomic data used in the theoretical models.  
Different atomic datasets are available for different emission-lines, and these may introduce errors in the estimated quantities.  \citet{Copetti02} compared different electron density indicators in planetary nebulae and concluded that atomic data was the cause of differences in \OII\ densities when compared with other indicators.   In this work, we include the most recent atomic data available, but we are still constrained to the accuracy of the atomic data and the fact that this data comes from different sources for different species.  We refer the reader to \citet{Proxauf14} for a detailed comparison and discussion of atomic datasets on electron density estimates.

\subsection{Metallicity Calibration Discrepancies}

The electron temperature in a nebula is fundamentally linked to the gas-phase metallicity because metals act as coolants in the nebula.  Therefore, many of the ISM pressure diagnostics have a strong dependence on the gas-phase metallicity.  Different metallicity calibrations exhibit extremely large systematic discrepancies (up to 1 dex in \OH) \citep[see][for a review and discussion]{Kewley08}.  Due to these discrepancies, our ISM pressure calibrations should only be used with metallicity calibrations that have been constructed using consistent theoretical models (i.e with Starburst99 and MAPPINGS v5 photoionization models), such as the metallicity surface fits in Kewley et al. (ARAA, submitted), or the code NebulaBayes, which calculates the gas-phase metallicity of both star-forming and AGN regions using MAPPINGS v5 models \citep{Thomas18}.

\subsection{Diffuse Ionized Gas}

The ionizing radiation produced by the star clusters is partially absorbed by the \HII\ regions.  Superbubble models suggest that the combined effects of supernovae, stellar winds, and large-scale ionization by OB associations create a complex density and ionization structure that can be porous to ionizing radiation, allowing some radiation to escape \citep{Shields90}.  The absorption by the \HII\ region produces a hard ionizing radiation field incident on gas outside \HII\ regions in the disk, as well as on gas up to several kpc above the disk.  The gas that receives ionizing radiation that has leaked from nearby \HII\ regions is referred to as Diffuse Ionized Gas (DIG).    

The radiation in the DIG may come from a variety of sources.  \citet{Martin97} showed that the DIG has a radial gradient that is consistent with the dilution of radiation from a centralized source, indicating that the dominant excitation mechanism of the DIG is photoionization by the radiation from massive stars.  However, in some galaxies, 30-50\% of the DIG emission may come from shock excitation \citep{Martin97,Ramirez-Ballinas14}, and a very minor component ($<20$\% of \Ha) may come from dust scattered radiation \citep{Barnes15,Ascasibar16}.  The DIG can also be produced by evolved stars \citep{Zhang17}.  Hot evolved stars, such as post-AGB stars, are characterized by very high temperatures.  Zhang et al. used photoionization models to show that high temperature evolved stars can produce a hard ionizing spectrum.  By comparing their models with spatially resolved data of nearby galaxies, they conclude that hot evolved stars make a major contribution to the DIG.

The DIG has a different emission-line spectrum to star-forming regions.  The escaped hard radiation field produces strong emission in low-ionization species, such as \OI, \NII, \CII, and \SII, but weak emission in high ionization species such as \OIII\ \citep[see reviews by][]{Rand98,Mathis00,Haffner09}.   The DIG may contribute between 10-50\% of the \Ha\ emission, but may contaminate the \NII, \SII\ and other low ionization lines up to 2-3$\times$ more \citep{Madsen06,Oey07,Blanc09}.  

Theoretical models are needed to help remove the DIG component from spectra.  Our photoionization models were calculated assuming a normal star forming source, and do not include a hard component that would be typical of the DIG.  
Current models can only partially reproduce the DIG line ratios \citep[e.g.,][]{Bland-Hawthorn97}. For example, 
\citet{Martin96} used a combination of photo-ionized gas with shock-ionized gas using various shock speeds, and turbulent mixing layers of various temperatures to model the DIG in irregular galaxies.  However, a fully self-consistent model of the effect of the DIG on the UV-IR emission-lines is still lacking.  

Care must be taken to account for the DIG when using global spectra of galaxies, or when using integral field data where \HII\ regions are not spatially resolved.  The pressure or density derived using low ionization lines may not represent a luminosity-weighted average \HII\ region within a galaxy.  Instead, the pressure or density estimates may include a component from the pressure or density within the DIG, depending on the surface brightness of the DIG, and the sensitivity of the observations.  We are currently investigating the effect of the DIG on multiple emission-line diagnostics, including our pressure and density diagnostics (Poetrodjojo et al., in prep). With such information, DIG models might allow reliable ISM pressures to be calculated for the global spectra of galaxies, taking into account possible DIG contamination.  

\subsection{Shock excitation}

Shock excitation can be produced by many phenomena, including supernovae, stellar winds, galaxy interactions, and AGN related activity such as jets.  The effect of shocks on the optical emission-lines has been observed in both interacting and isolated galaxies with high spectral resolution integral field spectroscopy \citep{Ho14,Rich15,Dopita15,Ho16}.  Gas outflows and mergers can produce widespread shocks throughout galaxies which significantly affect the emission-line spectrum of a galaxy at both kpc-scales and sub-kpc scales within a galaxy \citep{Medling15}.

Shock excitation can contaminate both low ionization line ratios and high ionization line ratios, depending on the shock velocity.  Fast shocks ($v>500\, {\rm km s}^{-1}$) produce strong high ionization lines such as \OIII\ \citep{Allen98,Allen08}, while slow shocks produce relatively weak high ionization lines, but strong low ionization species such as \SII\ and \NII\ \citep{Rich11,Rich15}.  This difference occurs because fast shocks produce a photoionizing precursor which produces significant ionization in front of the shock.

Shocks can be identified using a combination of morphological information, velocity maps, velocity dispersion, and emission-line ratios \citep[see][for examples of how shocks can be identified at low and high redshift]{Ho14,Yuan12}.  We recommend that these tools be used to identify galaxies containing shocks prior to the application of the pressure and density diagnostics given here, because our pressure and density diagnostics have been produced using the ionizing radiation from pure stellar populations, not the harder radiation field from shocks, which changes the ionizing spectrum, and may change the electron density distribution.  For example, shocked regions from galactic outflows can have higher electron densities, as seen in edge-on disk galaxies from the SAMI galaxy survey \citep{Ho14}.  

High resolution integral field spectroscopy allows the shock component to be resolved and separated from the \HII\ region component to optical emission-line spectra by identifying emission-lines with different velocity dispersions (e.g., \citet{Ho14}, D'Agostino et al., in prep).  In these cases, it is feasible to apply our pressure and density diagnostics to the \HII\ region component of the spectra.  For galaxies where shocks have been identified, but the shock and \HII\ region components cannot be reliably separated, our pressure and density diagnostics should not be used because the diagnostics may yield unreliable densities and pressures.

\subsection{Measuring ISM pressure and electron density in global spectra}

Understanding the ISM pressures derived for global spectra of galaxies (or for large spaxels within galaxies where \HII\ regions are unresolved) is non-trivial.  The main complication with measuring the electron density or ISM pressure for global spectra is saturation.  If the emission-line ratios are saturated (i.e. in the low density limit or the high density limit) in some, but not all regions of the galaxy, the resulting density or pressure estimate will not represent the luminosity-weighted average density or pressure.  In this case, the line ratios will be strongly affected by the fraction of regions that are in the low density limit and the fraction of regions that are in the high density limit.  To overcome this problem, several line ratios that cover a range of critical densities could be used together to identify which line ratios (if any) can sample the full set of pressures across the galaxy without being saturated.

If a line ratio, or a set of line ratios have been identified that sample the full set of pressures across the galaxy with no saturation, the resulting global pressure will represent the ISM pressure within a luminosity-weighted equivalent \HII\ region.  In the global spectrum of a galaxy, the electron density derived from our calibrations is the density that exists within an equivalent luminosity-weighted \HII\ region with a temperature structure that is the same as our stellar photoionization model at the model metallicity and density across the \HII\ region.   

For ultracompact \HII\ regions or \HII\ regions with complex geometries, this constant density assumption is likely to be invalid.  In these cases, our density diagnostics would only provide an average estimate of the density, at best.  We recommend the use of tailored \HII\ region models for ultracompact \HII\ regions for these cases, such as in \citet{Dopita06b}.

Modeling how well the integrated properties of an entire galaxy can be calculated using these diagnostics is significantly more complicated.  Realistic simulations of observed integrated spectra requires modeling of the star formation, metallicity, pressure, dust, and ionization parameter distributions within galaxies, weighting by surface brightness to simulate integrated data at different redshifts, and convolving the output spectrum through given instrument response functions.  

\section{Conclusions}\label{conclusions}

We have used new self-consistent stellar evolution and photoionization models to produce new ISM pressure and density diagnostics from the UV to the far-infrared.  
Our ISM pressure diagnostics take into account the temperature and density structure of the nebula, while our electron density diagnostics take the temperature structure into account and assume a constant density.  We anticipate that these diagnostics will provide a useful probe of the pressure and density within \HII\ regions and galaxies.

We show that our pressure diagnostics primarily depend on the gas-phase metallicity (through its effect on the electron temperature of the gas), with only a minor effect from the ionization parameter.  

We show how different species probe different zones within a nebula.  We anticipate that measuring multiple lines with different ionization states and different species will provide an unprecedented a picture of the detailed pressure or density structure within \HII\ regions.  

We show that the \SII\ and \OII\ lines do not probe similar regions of a nebula and that they are likely to only yield comparable results under constant electron temperature and density conditions.  The UV \NII\ and optical \OII\ lines probe similar nebular zones, and may be used to compare the ISM pressure or electron density in galaxies across cosmic time.

In the near future, {\it JWST} and the ELTs will enable the measurement of multiple ISM pressure-sensitive line ratios for galaxies across cosmic time, providing an important probe of how the luminosity-weighted average pressure in galaxies has changed over time.  

\acknowledgments

We wish to thank the referee for his/her extensive and insightful comments which have significantly improved this paper.  Parts of this research were supported by the Australian Research Council Centre of Excellence for All Sky Astrophysics in 3 Dimensions (ASTRO 3D), through project number CE170100013.  L.J.K. gratefully acknowledges the support of an ARC Laureate Fellowship (FL150100113).  This research has made use of NASA's Astrophysics Data System Bibliographic Services.


\newpage
\LongTables
\begin{landscape}
\tablecolumns{16} 
\begin{deluxetable*}{llllllllllllllll}
\tabletypesize{\footnotesize}
\tablecaption{Theoretical flux ratios for our MAPPINGS v5.1 Pressure Models \label{pressure_flux_table}}
\tablehead{  
  \colhead{log} & 
  \colhead{log(q)} &   
  \colhead{log(O/H)} &  
  \colhead{\NIV} & 
  \colhead{SiIII]}  & 
  \colhead{\CIII} & 
  \colhead{\NeIV} &  
  \colhead{\AlII}  & 
  \colhead{\OII}  &  
  \colhead{\ArIV} & 
  \colhead{\NI} & 
  \colhead{\ClIII} & 
  \colhead{\SII} & 
  \colhead{\SIII} & 
  \colhead{\OIII} & 
  \colhead{\NII} \\
   \colhead{($\frac{P}{k}$)} &
    \colhead{} & 
    \colhead{+12} & 
    \colhead{$\frac{\lambda 1483}{\lambda 1486}$} & 
    \colhead{$\frac{\lambda 1883}{\lambda 1892}$} & 
    \colhead{$\frac{\lambda 1906}{\lambda 1908}$} & 
    \colhead{$\frac{\lambda 2425}{\lambda 2423}$} & 
    \colhead{$\frac{\lambda 2660}{\lambda 2669}$} & 
    \colhead{$\frac{\lambda 3729}{\lambda 3726}$} &
    \colhead{$\frac{\lambda 4711}{\lambda 4740}$} & 
    \colhead{$\frac{\lambda 5198}{\lambda 5200}$} & 
    \colhead{$\frac{\lambda 5517}{\lambda 5537}$} & 
    \colhead{$\frac{\lambda 6717}{\lambda 6731}$} &
    \colhead{$\frac{\lambda 33 \mu m}{\lambda 18 \mu m}$} & 
    \colhead{$\frac{\lambda 52 \mu m}{\lambda 88 \mu m}$} & 
    \colhead{$\frac{\lambda 205 \mu m}{\lambda 122 \mu m}$} } 
\startdata
4.0 & 6.50 & 7.63 & \nodata & 1.5647 & 1.4862 & \nodata & 1.5352 & 1.4533 & \nodata & 0.69806 & 1.4216 & 1.4460 & 1.7769 & 1.5924 & 1.7000 \\
4.0 & 6.50 & 8.23 & \nodata & 1.5638 & 1.4813 & \nodata & 1.5331 & 1.4537 & \nodata & 0.69670 & 1.4263 & 1.4478 & 1.7929 & 1.6182 & 1.7301 \\
4.0 & 6.50 & 8.53 & \nodata & 1.5622 & 1.4728 & \nodata & 1.5290 & 1.4544 & \nodata & 0.69366 & 1.4341 & 1.4514 & 1.8192 & 1.6712 & 1.7865 \\
4.0 & 6.50 & 8.93 & \nodata & 1.5513 & 1.4479 & \nodata & 1.5117 & 1.4573 & \nodata & 0.68227 & 1.4482 & 1.4637 & 1.8778 & 1.8575 & 1.9474 \\
4.0 & 6.50 & 9.23 & \nodata & \nodata & \nodata & \nodata & \nodata & 1.4618 & \nodata & 0.66746 & 1.4398 & 1.4633 & 1.9378 & 1.9830 & 2.1429 \\
4.0 & 6.75 & 7.63 & \nodata & 1.5654 & 1.4898 & \nodata & 1.5370 & 1.4531 & \nodata & 0.69893 & 1.4177 & 1.4444 & 1.7648 & 1.5707 & 1.6761 \\
4.0 & 6.75 & 8.23 & \nodata & 1.5642 & 1.4837 & \nodata & 1.5344 & 1.4535 & 1.3706 & 0.69749 & 1.4239 & 1.4465 & 1.7855 & 1.6011 & 1.7130 \\
4.0 & 6.75 & 8.53 & \nodata & 1.5624 & 1.4738 & \nodata & 1.5297 & 1.4543 & 1.3842 & 0.69435 & 1.4331 & 1.4506 & 1.8165 & 1.6604 & 1.7783 \\
\enddata
\end{deluxetable*}
\clearpage
\end{landscape}

\clearpage 
\LongTables
\begin{landscape}
\tablecolumns{16} 
\begin{deluxetable*}{llllllllllllllll}
\tabletypesize{\footnotesize}
\tablecaption{Theoretical flux ratios for our MAPPINGS v5.1 Density Models \label{density_flux_table}}
\tablehead{  
  \colhead{log($n_e$)} & 
  \colhead{log(q)} &   
  \colhead{log(O/H)} &  
  \colhead{\NIV} & 
  \colhead{SiIII]}  & 
  \colhead{\CIII} & 
  \colhead{\NeIV} &  
  \colhead{\AlII}  & 
  \colhead{\OII}  &  
  \colhead{\ArIV} & 
  \colhead{\NI} & 
  \colhead{\ClIII} & 
  \colhead{\SII} & 
  \colhead{\SIII} & 
  \colhead{\OIII} & 
  \colhead{\NII} \\
   \colhead{} &
    \colhead{} & 
    \colhead{+12} & 
    \colhead{$\frac{\lambda 1483}{\lambda 1486}$} & 
    \colhead{$\frac{\lambda 1883}{\lambda 1892}$} & 
    \colhead{$\frac{\lambda 1906}{\lambda 1908}$} & 
    \colhead{$\frac{\lambda 2425}{\lambda 2423}$} & 
    \colhead{$\frac{\lambda 2660}{\lambda 2669}$} & 
    \colhead{$\frac{\lambda 3729}{\lambda 3726}$} &
    \colhead{$\frac{\lambda 4711}{\lambda 4740}$} & 
    \colhead{$\frac{\lambda 5198}{\lambda 5200}$} & 
    \colhead{$\frac{\lambda 5517}{\lambda 5537}$} & 
    \colhead{$\frac{\lambda 6717}{\lambda 6731}$} &
    \colhead{$\frac{\lambda 33 \mu m}{\lambda 18 \mu m}$} & 
    \colhead{$\frac{\lambda 52 \mu m}{\lambda 88 \mu m}$} & 
    \colhead{$\frac{\lambda 205 \mu m}{\lambda 122 \mu m}$} } 
\startdata
0.0 & 6.50 & 7.63 & \nodata & 1.5648 & 1.4863 & \nodata & 1.5355 & 1.4524 & \nodata & 0.69952 & 1.4211 & 1.4443 & 1.7721 & 1.5437 & 1.6956 \\
0.0 & 6.50 & 8.23 & \nodata & 1.5638 & 1.4814 & \nodata & 1.5333 & 1.4528 & \nodata & 0.69787 & 1.4259 & 1.4463 & 1.7882 & 1.5701 & 1.7266 \\
0.0 & 6.50 & 8.53 & \nodata & 1.5622 & 1.4729 & \nodata & 1.5291 & 1.4536 & \nodata & 0.69439 & 1.4338 & 1.4503 & 1.8147 & 1.6240 & 1.7837 \\
0.0 & 6.50 & 8.93 & \nodata & 1.5514 & 1.4480 & \nodata & 1.5120 & 1.4568 & \nodata & 0.68253 & 1.4481 & 1.4631 & 1.8748 & 1.8250 & 1.9453 \\
0.0 & 6.50 & 9.23 & \nodata & \nodata & \nodata & \nodata & \nodata & 1.4634 & \nodata & 0.66730 & 1.4394 & 1.4642 & 1.9449 & 2.1001 & 2.1523 \\
0.0 & 6.75 & 7.63 & \nodata & 1.5654 & 1.4899 & \nodata & 1.5373 & 1.4521 & \nodata & 0.70067 & 1.4172 & 1.4426 & 1.7598 & 1.5200 & 1.6716 \\
0.0 & 6.75 & 8.23 & \nodata & 1.5642 & 1.4837 & \nodata & 1.5346 & 1.4526 & 1.3704 & 0.69878 & 1.4235 & 1.4451 & 1.7806 & 1.5510 & 1.7097 \\
0.0 & 6.75 & 8.53 & \nodata & 1.5624 & 1.4739 & \nodata & 1.5297 & 1.4534 & 1.3841 & 0.69512 & 1.4328 & 1.4495 & 1.8119 & 1.6110 & 1.7760 \\
\enddata
\end{deluxetable*}
\clearpage
\end{landscape}

\clearpage

\begin{table}[htbp]
  \centering
 \caption{Density-sensitive lines and their ionization potentials}\label{table_ne}
     \begin{tabular}{rrrrrrr}
   \\  \hline
    Ion   & Line ratio & Ionization & min n$_e$\footnotemark[a] & max n$_e$\footnotemark[a] & $\Delta$n$_e$\footnotemark[b] & Case\footnotemark[c] \\
          &       &  potential (eV)    &       &       &  &  \\
    \hline
    UV    &       &       &       &       &       &  \\
    \hline
    \rion{S}{5}    & 1188/1199 & 47.222 & 60000 & 80000000 & 88\%  & 2 \\
    \rion{O}{5}     & 1214/1218 & 77.414 & 8000  & 4000000 & 41\%  & 2 \\
    \rion{N}{4}    & 1483/1486 & 47.445 & 2500  & 2500000 & 93\%  & 2 \\
    \rion{Si}{3}  & 1883/1892 & 16.346 & 900   & 450000 & 75\%  & 2 \\
    \rion{C}{3}  & 1906/1909 & 24.385 & 1800  & 630000 & 39\%  & 2 \\
    \rion{Ne}{4}   & 2423/2425 & 63.423 & 2500  & 660000 & 23\%  & 1 \\
    \rion{Al}{2}   & 2661/2670 & 5.986 & 450   & 180000 & 52\%  & 2 \\
    \hline
    Optical &       &       &       &       &       &  \\
    \hline
     \rion{O}{2}   &3727/3729 & 13.618 & 40    & 10000 & 59\%  & 1 \\
    \rion{Ar}{4}   &  4711/4740 & 40.735 & 350   & 80000 & 40\%  & 1 \\
   \rion{N}{1}     & 5200/5198 & 0.000     & 200   & 6300  & 3.5\% & 1 \\
    \rion{Cl}{3} & 5517/5537 & 23.813 & 250   & 50000 & 31\%  & 1 \\
   \rion{S}{2}   & 6717/6731 & 10.360 & 40    & 5000  & 47\%  & 1 \\
      \hline
    IR    &       &       &       &       &       &  \\
    \hline
  \rion{S}{3}  & 335$\mu$/187$\mu$ & 23.338 & 400   & 12000 & 14\%  & 3 \\
     \rion{O}{3}   & 88$\mu$/52$\mu$ & 35.121 & 60    & 3000  & 26\%  & 3 \\
    \rion{N}{2}    & 205$\mu$/122$\mu$ & 14.534 & 10    & 200   & 18\%  & 3 \\
    \hline
\end{tabular}
\vspace{5ex}

\footnotetext[a]{Minimum and maximum values of n$_e$ for which the line ratio is density sensitive.}

\footnotetext[b]{The uncertainty in $n_e$ due to the electron temperature, defined as the difference between the n$_e$ at log(T$_e$)=3.5 and 4.5, divided by 2$\times$ the value at log(T$_e$) at 4.0, at the mid-point of the values of the flux ratio over which the ratio is density sensitive.}

\footnotetext[c]{The case for which electron density sensitivity can arise in line ratios, as described in Section~\ref{density_temperature}.}
\end{table}

\end{document}